\documentclass[12pt]{article}
\usepackage[cp1251]{inputenc}
\newcommand{\di}{\displaystyle}

\fussy
\begin{document}

{\bf Three New Models of Layered Inhomogeneous Elliptical Galaxies}
\vskip 2ex
S. A. Gasanov
\vskip 1.5ex
Sternberg Astronomical Institute, Lomonosov Moscow State University, Moscow, 119991 Russia
e-mail: gasanovsa57@gmail.com
\vskip 1.5ex

{\bf Abstract.} To solve some problems of celestial mechanics and astrophysics, three new models of an elliptical
galaxy (EG) have been created, which are in good agreement with modern understanding of the structure of
such galaxies. Based on these models, the total gravitational (potential) energy and rotational kinetic energy
of an EG, as well as the velocity dispersion at a distance of its effective radius, are determined. A new method
is proposed for determining the average values of the scale radius of an EG and the density at its center, as
well as the average value of its key parameter  density $\beta$ and its value at a distance corresponding to the
effective radius of the galaxy. The results obtained are applied to sixty EGs and presented in the form of tables
for ten galaxies.
\vskip 1.5ex


\vskip 2ex

1. INTRODUCTION
\vskip 2ex

In [1, 2], the problem of the spatial motion of a passively gravitating body (a star or the center of mass of a
globular cluster, GC) inside a rotating elliptical galaxy
(EG) was considered. An EG was regarded as a two - layer ellipsoidal body: its luminous part is a triaxial
ellipsoid, and the space between the boundaries of its
luminous part and the halo is a homeoid filled with
homogeneous dark matter.

In [1], the luminous part of an EG (LP EG) is considered a homogeneous triaxial ellipsoid, and, in [2],
uses the so-called astrophysical law of density distribution. The EG model considered in [1] will be called
model 1, and the model in [2], model 2. Within models 1 and 2, an analogue of the Jacobi integral was
found and the region of possible motion of the star (or
the mass center of the GC) was determined. The type
and Lyapunov stability of the stationary solutions 
found libration points — were established, and zero 
velocity surfaces were constructed.

The problem of the spatial motion of a star inside
(near) a globular cluster belonging to an inhomoge
neous rotating EG was considered in [3]. The motion
of a star near a GC was considered with allowance for
the perturbations caused by the attraction of the EG,
which, together with the halo, is a two-layer body [1].
The motion of a star near a GC takes place outside the
luminous part of the EG, but inside the homeoid. The
notion "near a GC" as a "sphere of action" (Hill’s
gravitational sphere) has been refined and concretized. In connection with the introduced concept of the
sphere of action, two variants of the motion of a star
were considered: inside and outside the sphere of
action of a globular cluster, and the regions of the possibility of motion were determined. A quasi - integral
and minimum - energy surfaces were found, which,
under certain conditions, are transformed into an analog of the Jacobi integral and zero-velocity surfaces,
respectively.

The results of [1–3] are applied to model elliptical
galaxies with parameters that exactly coincide with the
parameters of the elliptical galaxies NGC 4472 (M
49), NGC 4636, and NGC 4374 (M 84) and are presented in the form of figures and tables.
When finding libration points and analyzing their
stability, exact expressions for the potentials of the LP
EG and the homeoid rather that their series expansions are used.

In this paper, three new models of an EG are considered, in which a galaxy with a halo is considered a
two-layer inhomogeneous ellipsoid of revolution: a
spheroid. In this case, the outer and inner layers are
assumed to be similar and concentric, and their centers coincide with the center of the EG. The LP EG is
considered an inner layer and comprises an inhomogeneous ellipsoid of revolution (spheroid) with a
homothetic (spheroidal) density distribution, or a layered inhomogeneous spheroid.

In the LP EG, baryon mass (BM) with an “astrophysical law” of density distribution prevails. The
outer part is an inhomogeneous spherical layer with a
spherical density distribution (model 3) or a spheroidal layer (conditionally, a homeoid) with a spheroidal
density distribution (model 4). According to model 3, the outer layer and the halo of a galaxy are bounded by
a sphere with a radius equal to the scale radius of the
EG, and, according to model 4, they are bounded by a
spheroidal surface with a semi-major axis equal to the
scale radius of the galaxy. It is believed that the spherical layer and the conditional homeoid mainly consists
of dark matter (DM) and, depending on its presence in
the inner (central) regions of the EG, in models 3 and
4, two variants are considered. Variant (a), in which
the main part of DM is outside the LP EG [4], and
variant (b), in which the DM content in the inner
regions of the EG is comparable to the BM content [5,
6]. In this case, the matching conditions for the potentials at the interface between the LP EG and the 
spherical layer or conditional homeoid are determined.

As correctly noted in [4], the nature of DM is
unknown, and there is no clear understanding of its
physical relationship with the observed astronomical
objects. Nevertheless, its presence in galaxies is recognized and indirectly confirmed. In this work, three
types of EG together with a halo are simulated, which
cannot claim to be comprehensive of the DM problem
as a whole.

According to model 5, an EG with (variant 1) or
without (variant 2) a halo comprises a layered inhomogeneous spheroid consisting of BM and DM. In
model 5, there is no interface between the LP EG and
the conditional homeoid; therefore, the fulfillment of
the matching conditions for the potentials is not considered. In models 3–5, the conditional boundaries
(apparent sizes) of the LP EG are determined by the
quantities  $D_{25}$  and  $R_{25}$  \ [7].

Beside the problem of the presence or absence of
DM in the inner regions of an EG, there is another
problem, namely, the true spatial orientation of the
galaxy, which is unknown to us. Finding the true shape
of such galaxies is of great importance for constructing
a dynamic theory of equilibrium and the theory of the
origin of EGs. As shown in [8], the true shape of an
EG can be determined on the basis of two observational tests of 1) whether the rotation axis coincides
with the apparent minor axis of the galaxy and 2)
whether the isophotes of the galaxy are aligned. The
use of these tests allows one to judge more confidently
about the shape of EGs and indicates the existence of
EGs both in the form of oblate or prolate spheroids
and triaxial ellipsoids. If an EG has the shape of a prolate spheroid or a triaxial ellipsoid, then the rotation
axis will, as a rule, be observed as not aligned with the
apparent minor axis and the isophotes of the galaxy
will be misaligned. If the galaxy has the shape of an
oblate spheroid, then the projection of the minor axis
onto the plane of the sky coincides with the direction
of the rotation axis for any orientation of the galaxy
relative to the observer and the isophote alignment is
not disturbed.

In [9], a new method for solving the inverse geometric problem of recovering the shape of ellipsoidal
bodies through their projection (limb) onto the plane
of the sky was developed and used. Using this method,
the semiaxes of the dwarf planet Haumea as a triaxial
ellipsoid were determined. Thus, for each value of the
photometric parameters of the dwarf planet, its shape
and average density were determined, as well as the
orientation relative to its ring and the orbits of satellites. It turned out that the planes of the planet’s ring
and the orbit of its satellite Hi’iaka do not coincide
with the planet’s equator plane and both satellites are
in direct motion [9].

It is important to note that the method developed
in [9] and applied to the dwarf planet Haumea cannot
be used to recover the true shape of the EG as a triaxial
ellipsoidal body or spheroid. In the present work, an
EG is regarded as a layered inhomogeneous prolate
ellipsoid of revolution (or a prolate spheroid for short).
For a triaxial ellipsoid, the rotation axis does not coincide with the apparent minor axis and the isophotes
will be misaligned. A more convenient or simpler variant in which an EG is an oblate spheroid, with its
apparent axis coinciding with the axis of rotation and
the isophote alignment not disturbed, is not considered.

Within the created models, a new method is proposed for determining the average values of the EG
scale radius $r_s$ and the densities in the center, $\rho_0$, and
at the boundary of the galactic halo, $\rho_s$. The total gravitational (potential) energy $W$ and rotational kinetic
energy $T_{rot}$ of an inhomogeneous EG, the velocity dispersion  $\sigma_{eff}$ at a distance of the effective radius of the
galaxy $R_{eff}$ , as well as the average value of the parameter $\beta$  and its value $\beta_{eff}$  corresponding to the effective radius of the galaxy, are determined according to these
models.

Thus, it is currently not possible to accurately
determine the true shape of an EG as a triaxial ellipsoid or even a spheroid. Therefore, we determine the
values of the apparent major and minor semiaxes of
the EG. Thus, the obtained values of the dynamic
parameters listed above are approximate.

The problem of equilibrium and stability of
dynamical systems appearing in these models is of particular interest and will be considered in a separate
work.

\vskip 2ex
2. DENSITY DISTRIBUTION LAWS
\vskip 2ex

Assume that the luminous part of an EG (LP EG)
is a layered inhomogeneous ellipsoid of revolution
(prolate spheroid) with semiaxes \ $a > b = c$. As the
density distribution law (LP EG profile), consisting of
the BM, we take the “astrophysical” profile \ $\rho\,(m)$   [10, 11]
obtained by applying the Abel integral equation to
the Hubble surface brightness profile \ $I\,(m)$ [12]: 
$$
     \rho\,(m) = \frac{\rho_0}{\left(1 + \di \beta m^2\right)^{3/2}}, \quad  I\,(m) = \frac{I_0}{1 + \di \beta m^2}, 
\eqno (1) $$
where \ $\rho_0$  and  $I_0$ are the density at the center of the EG
and the central surface brightness and the parameter \ $\beta \gg 1$ 
 for each EG is selected separately and is found
by aligning the photometry data [10, 11]. In addition,
 $m$ is a parameter of a family of similar and concentric
spheroids
$$
   \frac{x^2}{a^2} +  \frac{y^2 + z^2}{c^2} = m^2,  \qquad (a > b = c, \quad  0 \le m \le 1),
\eqno (2) $$
Here, \ $m = 0$  corresponds to the center of the EG and
 \ $m = 1$ corresponds to the spheroidal surface limiting
the LP EG.

Next, we assume that an inhomogeneous spherical
layer (model 3) is filled with DM with a density distri-
bution law (profile) [13, 14]
$$
   \rho_S(r) = \frac{K r_s}{r} \left( 1 + \frac{r}{r_s}\right)^{-\,2}, \quad
    \widetilde \rho_S \equiv  \rho_S(r_s) = \frac{K}{4}
\eqno (3) $$
Here, \ $K$ is a normalizing coefficient, which has the
dimension of density in the solar masses per cubic parsec, $r_s$ is the scale radius of the EG, and 
  $\widetilde \rho_S$ is the density of DM at the outer boundary of the layer. Formulas for calculating the 
  coefficient $K$ are given in [3, 13,14].

It should be noted that, in addition to profile (3),
there are other laws of density distribution. In the book
[15], a more general profile is given, from which, as
special cases, the Dehnen [16, 17], Hernquist [18],
Jaffe [19], and NFW [13] profiles are obtained.
If an inhomogeneous spheroidal layer (model 4)
filled with DM is considered, then profile (3) cannot
be used. In this case, the density distribution law,
which we call an analog of the NFW profile, is
defined as
$$
   \rho_G(\mu) =  \frac{K}{\xi \mu ( 1 +  \xi \mu)^2},    \quad 
      \left( \frac{r}{r_s} = \xi\, \mu, \quad \xi = \frac{\di \sqrt[3]{\widetilde a \widetilde c^2}}{r_s} < 1\right), 
 \eqno (4) $$
which was proposed by B.P. Kondratyev. Here, $\mu$ is
the parameter of a family of similar and concentric
spheroids with semiaxes \ $\widetilde a$ and $\widetilde b = \widetilde c$: 
$$  
    \mu^2 =  \frac{x^2}{\widetilde a^2} +  \frac{y^2 + z^2}{\widetilde c^2}, 
    \quad (\widetilde a = r_s > \widetilde b = \widetilde c) 
 \eqno (5) $$
 In this case, families (2) and (5) are assumed to be
similar and concentric spheroids. The value $\mu^2 = 1$
corresponds to the outer boundary of the spheroidal
layer, at which, according to expression (4), the density is equal to
$$
   \bar \rho_G \equiv \rho_G(1) =  \frac{K}{\xi\,( 1 +  \xi)^2}
\eqno (6) $$

\vskip 4.0ex

3. POTENTIAL ENERGY AND ROTATIONAL 
KINETIC ENERGY OF THE LUMINOUS PART 
OF AN ELLIPTICAL GALAXY

\vskip 2.0ex

The total potential (gravitational) energy  $W\,(1)$ and
rotational kinetic energy  $T\,(1)$ of a layered inhomogeneous spheroid with density  and semiaxes
\ $a > b = c$\ are determined by the formulas [10]
$$
   W\,(1) = -\, \pi G\,a c^2 J_0\, \Psi\,(1), \quad    T\,(1) = \frac{\pi G a c^2}{2}\,J_1\Psi\,(1), 
   \quad J_1 = J_0 - 3c^2K_0
  \eqno (7) $$
respectively. Here, $\Psi\,(1)$ is the value of the function 
$$
  \Psi\,(m)  = \int \limits_0^{m^2} \rho\,(m)\, M\,(m)\,dm^2, \quad  
   M\,(m)  =  4 \pi  a c^2 \int \limits_0^m m^2\, \rho\,(m)\,dm
\eqno (8)$$
at $m = 1$, and 
$$
   J_0 = \int \limits_0^{\infty} \frac{du}{\Delta\,(u)}, \quad 
    K_0 =  \int \limits_0^{\infty} \frac{du}{\di (c^2 + u)\,\Delta\,(u)}, \quad \Delta\,(u) = (c^2 + u)\,\sqrt{a^2 + u},
\eqno (9) $$
or
$$
   J_0 =  \frac{1}{\di \sqrt{a^2 - c^2}}\di \ln \frac{a + \sqrt{a^2 - c^2}}{a - \sqrt{a^2 - c^2}}, \quad 
   K_0 = \frac{2a - c^2J_0}{2c^2(a^2 - c^2)}, \quad J_1 = \frac{(2a^2 + c^2)J_0 - 6a}{2(a^2 - c^2)}
\eqno (10) $$
If the density \ $\rho\,(m)$  of the LP EG is determined by
astrophysical law (1), then, by virtue of (8), the 
mass $M\,(m)$ of the intermediate ellipsoid and the function $\Psi\,(m)$ will be
$$
    M\,(m) = 4\pi \rho_0 a\,c^2 g_1(m), \quad   \Psi\,(m) = 4\pi \rho_0^2 a c^2 f_1(m)
\eqno (11)$$
Here
$$
  g_1(m) =  \frac{1}{\beta \, \sqrt{\beta}}\, \left[ \ln \varphi_1(m) - \frac{h}{\di \sqrt{1 + h^2}}\, \right], \quad 
 \varphi_1(m) = h + \sqrt{1 + h^2}, \quad h = \sqrt{\beta}\,m
\eqno (12) $$
$$  
     f_1(m) = \di \frac{1}{\beta^2 \sqrt{\beta}}\, \left[{\it arctg }\,h +
     \frac{h}{1 + h^2} -   \frac{2 \ln\,\varphi_1(m)}{\sqrt{1 + h^2}}  \right]
\eqno (13)$$
Then, by virtue of (7), we find the gravitational energy
  $W_L = W\,(1)$ of the LP EG and its rotational kinetic
energy \ $T_L = T\,(1)$, according to variant (a), in the form
$$
 a) \quad   W_L = -\, 2W_0J_0, \quad   T_L = W_0J_1, \quad  W_0 = 2\pi^2 G \rho_0^2 a^2c^4f_1(1)
\eqno (14) $$   
where the function $f_1(m)$ and coefficients \ $J_0$  and $J_1$  are
defined above.

In variant (b), the gravitational energy \ $\widetilde W_L$ of the LP
EG and its rotational kinetic energy \ $\widetilde T_L$ are determined
differently; namely, in formulas (7) and (8), the function  $\Psi\,(m)$ \ is replaced by \ $\widetilde \Psi\,(m)$. 
The latter is obtained
from expression (8) by replacing the profile   $\rho\,(m)$  and
mass  $M\,(m)$   by the general profile  $\rho\,(m) + \rho_G(m)$ and
the total mass $M\,(m) + \widetilde M_G(m)$ of BM and DM. Then,
the sought expressions $\widetilde W_L = \widetilde W_L(1)$ and $\widetilde T_L = \widetilde T_L(1)$  are
represented as (b)
$$
 b) \quad   \widetilde W_L = -\, 2\widetilde W_0J_0, \quad  \widetilde T_L = \widetilde W_0J_1, \quad 
 \widetilde W_0 = \frac{\pi G\,a c^2}{2}\, \widetilde \Psi\,(1), 
 \eqno (15) $$
where
 $$
  \widetilde \Psi\,(m) =  \sum_{n=1}^4\widetilde \psi_n(m),  
  \quad \widetilde \psi_2(m) =  \int \limits_0^{m^2}\,\rho\,(m)\, \widetilde M_G(m)\, dm^2, 
$$
$$
   \quad  \widetilde \psi_3(m) =  \int \limits_0^{m^2}\,\rho_G(m)\, M\,(m)\, dm^2, \quad
    \widetilde \psi_4(m) =  \int \limits_0^{m^2}\,\rho_G(m)\, \widetilde M_G(m)\, dm^2,  
$$
Here, $\widetilde \psi_1(m) \equiv \Psi\,(m)$, the function  $\Psi\,(m)$ is defined by
equality (11) and  is the mass $\widetilde M_G(m)$ of an inhomogeneous spheroid \ $a > b = c$  consisting of DM with a
profile $\rho_G(m)$  and is calculated by the formula
 $$
  \widetilde M_G(m) = 4 \pi  a c^2 \int \limits_0^m m^2\, \rho_G(m)\,dm
  = 4 \pi K\bar a^3 \left(\ln \bar g - \frac{\bar g - 1}{\bar g}\right),   \quad 
  \bar g = 1 + \bar \xi\, m, \quad \bar \xi = \frac{\sqrt[3]{a c^2}}{\bar a}
 \eqno (16) $$
Here, \ $\bar a$  is the scale of the galaxy. After calculating the
functions   \ $\widetilde \psi_k(m), \ (k = 2,3,4)$, we obtain for  $\widetilde \Psi\,(m)$ the following expression:
$$
   \widetilde \Psi\,(m) = 4\pi a c^2\left[\rho_0^2f_1(m) + 2K\rho_0 f_2(m) + K^2f_3(m)\right],
  \eqno (17) $$
where the function $f_1(m)$ is defined by equality (13), 
$$
  f_2(m) =  \frac{1}{\beta\,\bar \xi^3 \bar g\ }\left[ \frac{\bar g - 1}{\di \sqrt{1 + h^2}} - 
   \frac{\bar g - 1 }{h}\,\ln \varphi_1(m)  + \frac{\bar \xi \bar g }{2\sqrt{\beta + \bar \xi^2}}\, 
   \ln \frac{\varphi_2(m)}{\varphi_2(0)} -  \frac{\bar g}{\sqrt{1 + h^2}}\, \ln \bar g \right], 
  \eqno (18) $$
$$
f_3(m) =  \frac{1}{\bar \xi^5}\,\left(1 - \frac{1}{ \bar g^2} - \frac{2  \ln \bar g}{ \bar g}\right), \quad 
\varphi_2(m) = \di \frac{\beta\,m - \bar \xi + \sqrt{1 + h^2}\,\sqrt{\bar \xi^2 + \beta}}
   {-\,\beta\,m + \bar \xi + \sqrt{1 + h^2}\,\sqrt{\bar \xi^2 + \beta}},
 \eqno (19) $$
where the function $\varphi_1(m)$ and parameter  $h$  are defined
by equality (12) and  $\bar g$  and $\bar \xi$, by equality (16).

It is obvious that expression (15) for energies $\widetilde W_L$
and $\widetilde T_L$ coincides with expression (14) for energies $W_L$
and  $T_L$ if the LP EG consists only of BM. Indeed, in
this case $\rho_G(m) = 0$, or  $\widetilde \psi_k(m) = 0, \ (k = 2,3,4)$. Therefore,  
$\widetilde \Psi\,(m) = \Psi\,(m)$ and  $\widetilde W_0 = W_0$, or, which is the same,
 $\widetilde W_L = W_L$  and $\widetilde T_L = T_L$.  
 
It is important to note that the gravitational energy
and rotational kinetic energy of the EG corresponding
to variant (2) of model 5 are also determined by equality (15). In addition, from this equality, expressions for
these energies corresponding to variant (1) of model 5
can be easily obtained by replacing  \ $a, b = c$   with
 $\widetilde a, \widetilde b = \widetilde c$ and  $\bar a$  with  $r_s$.

\vskip 4 ex

 4. POTENTIAL ENERGIES AND ROTATIONAL 
KINETIC ENERGIES OF INHOMOGENEOUS 
SPHEROIDAL AND SPHERICAL LAYERS

\vskip 2 ex

First, let us determine the potential energy $W_G$ and
rotational kinetic energy $T_G$ of an inhomogeneous
spheroidal layer consisting of DM with a profile $\rho_G(\mu)$ 
and mass $M_G(\mu)$. We assume that the inner and outer
boundaries of this layer coincide with the spheroids
with semiaxes \  $\mu \, \widetilde a$, $\mu \,\widetilde b = \mu \, \widetilde c$  and 
$\widetilde a$, $\widetilde b = \widetilde c$, respectively.
Then, the energies $W_G$  and $T_G$  will be determined by
the formulas [11]
$$
  W_G = -\, \pi G  \widetilde a \widetilde c^2  J_0 \Psi_G(\mu),  \quad 
    T_G =   \frac{\pi G  \widetilde a \widetilde c^2}{2}J_1\Psi_G(\mu),
 \eqno (20)$$
 where the coefficients  $J_0$ and  $J_1$  are given above and
the function $\Psi_G(\mu)$ is defined by the formula
$$
    \Psi_G(\mu) = \int \limits_{\mu^2}^1 \rho_G(\mu) M_G(\mu) d\mu^2, \quad 
   M_G(\mu) =  4 \pi  \widetilde a \widetilde c^2 \int \limits_{\mu}^1 \mu^2\, \rho_G(\mu)\,d\mu
$$
After calculating the integrals for the mass of the
homeoid $M_G(\mu)$ and the function  $\Psi_G(\mu)$, we obtain
the following expressions:
$$
    M_G(\mu)  =  4 \pi K r_s^3 \left[\ln\,\frac{1 + \xi}{1 + \xi \mu} - \frac{\xi\,(1 - \mu)}{(1 + \xi)(1 + \xi \mu)}\right],
    \quad \Psi_G(\mu) = \frac{8 \pi K^2 r_s^3}{\xi^2}\,H_G(\mu),  
 \eqno (21) $$
where 
 $$
     H_G(\mu) = \frac{1}{1 + \xi \mu}\, \left[\frac{\xi (1 - \mu)(2 + 3 \xi + \xi \mu + 2\xi^2 \mu)}{2(1 + \xi)^2(1 + \xi \mu)} - \ln \di \frac{1 + \xi}{1 + \xi \mu}\right], \quad \xi = \frac{\sqrt[3]{\widetilde a \widetilde c^2}}{r_s}
$$
By virtue of (21), the energies  $W_G$  and $T_G$   can be
rewritten in a more compact form:
$$
  W_G = -\, 2\widetilde  W_0J_0H_G(\mu), \quad  T_G =  \widetilde  W_0J_1H_G(\mu),\quad 
   \widetilde  W_0 = 4 \pi^2GK^2\xi r_s^6
 \eqno (22)$$
Now let us calculate the potential energy $W_S(r)$  of an
inhomogeneous spherical layer (model 3) with a profile  $\rho_S(r)$ defined by equality (3), using the formula
from [11]:
$$
  W_S(r) =  -\,2 \pi  \int \limits_a^r u^2  \rho_S(u) \, U_S(u) du , \quad 
  U_S(r) = \frac{4 \pi G}{r} \, \int \limits_a^r u^2 \rho_S(u) du  + 4 \pi G \int \limits_r^{r_s} u\, \rho_S(u) du
$$

Next, we will consider that the radius of the inner cir-
cle of this layer is equal to $a$: the semi-major axis of the
midrange EG, and the radius of the outer circle is
equal to the scale radius $r_s$ of the EG. Then the internal
potential $U_S$ and mass $M_S(r)$ of the intermediate
spherical layer will be equal to
$$
    U_S(r) =  \di \frac{4 \pi G K r_s^3}{r}\,\left( \ln \frac{r + r_s}{a + r_s} -  \frac{r}{2r_s}
    + \frac{a}{a + r_s}\right), \    M_S(r) =  4 \pi K r_s^3
   \left(\ln \frac{r + r_s}{a + r_s} - \frac{r_s}{a + r_s} + \frac{r_s}{r + r_s} \right)
$$ 
respectively, and, at  $r = r_s$, we obtain the total mass of
this layer.

Let us substitute the expressions for the potential $U_S(r)$
 of this layer and its density $\rho_S(r)$ from (3) into
the expression for $ W_S(r)$. Then, for the total gravitational energy $W_S = W_S(r_s)$  of an inhomogeneous
spherical layer, we obtain
$$
  W_S = -\,8 \pi^2 G K^2r_s^5 \psi_1(r_s), \quad 
 \psi_1(r_s) = \ln \frac{a + r_s}{2r_s} +  \frac{(r_s - a)(5a + 3r_s)}{4 (a + r_s)^2} 
 \eqno (23) $$
The rotational kinetic energy $T_S(r)$ of a spherical layer
can be represented as the difference between the energies 
 $\widetilde E_1$ and $\widetilde E_2$  of two inhomogeneous balls of radii $r_s$
and  $a$ with the same density $\rho_S(r)$, respectively. To
determine the rotational energies $\widetilde E_1$ and $\widetilde E_2$, we can
use the formula
$$
  \widetilde E_k = \frac{1}{2}\,\widetilde J_k\Omega_k^2, \quad   \Omega_k^2 = \frac{G M_k}{r_k^3}, \quad (k = 1, 2),
$$
where $M_1 = M_H$ and $M_2 = M^*$ are the masses of the
halo and the LP EG, $r_1 = r_s$   and $r_2 = a$   are the scale
radius of the EG and the semi-major axis of its luminous part, and  and  are the moments of inertia of
inhomogeneous balls with density  $\rho_S(r)$ and radii  $r_s$ 
and  relative to the axis of rotation. To determine $\widetilde J_1$
and $\widetilde J_2$, we use the general formula for the moment
of inertia:
$$
 J\,(r) =  \frac{8 \pi}{3}\int \limits_0^r u^4 \rho_S(u) du
 = \frac{8\pi}{3} K\,r_s^3 \left[ 3r_s^2 \ln \frac{r + r_s}{r_s} +  \frac{r(r^2 - 3rr_s - 6r_s^2)}{2(r + r_s)} \right],  
$$

setting $r = r_s$  and $r = a$ and find, respectively,  $J_1$  and  $J_2$ 
 and, hence,  $\widetilde E_1$ and $\widetilde E_2$. Therefore, the sought for
expression for the rotational energy $T_S$ of a spherical
layer is obtained in the form
$$
 T_S = \widetilde E_1 - \widetilde E_2, \qquad \widetilde E_1 =  \frac{4\pi}{3}\,G M_H K\,r_s^2(3 \ln 2 - 2),
 \eqno (24)$$
$$
  \widetilde E_2 =  \frac{4 \pi GM^*}{3a^3}\,K\,r_s^3
   \left[ 3r_s^2 \ln \frac{a + r_s}{r_s} + \frac{a(a^2 - 3ar_s - 6r_s^2)}{2(a + r_s)} \right]
$$
  
\vskip 4 ex

5. TOTAL POTENTIAL AND KINETIC ENERGY OF A LAYERED INHOMOGENEOUS ELLIPTICAL GALAXY
\vskip 2ex

The total gravitational energy $W$  and rotational
kinetic energy $T$ of a dynamical system according to
variant (a) of models 3 and 4 will be presented in the
form of sums  $W = W_1 + W_2 + W_3$ and $T = T_1 + T_2$,
and, according to variant (b), in the form of sums
 $\widetilde W = \widetilde W_1 + W_2 + \widetilde W_3$ and $T = \widetilde T_1 + T_2$, respectively.
Here,$W_1, T_1$   and $\widetilde W_1, \widetilde T_1$ are the gravitational energies
and rotational energies of the LP EG in variants (a)
and (b), respectively. In addition, $W_2$ and $T_2$ are the
gravitational energy and rotational energy of the inhomogeneous spherical (spheroidal) layer $W_3$ and 
$\widetilde W_3$  are the mutual gravitational energies of the LP EG
and this layer according to variants (a) and (b), respectively. According to variant (a) of both models, the
energies $W_1 = W_L$ and $T_1 = T_L$  of the LP EG were
determined by equalities (14), respectively. In this
case, the energies $W_2 = W_G$   and $T_2 = T_G$ of the
homeoid are determined by equality (22) and the
energies $W_2 = W_S$ and $T_2 = T_S$ of the spherical layer
are determined by equalities (23) and (24), respectively.

The energies $\widetilde W_1 = \widetilde W_L$ and $\widetilde T_1 = \widetilde T_L$ corresponding
to variant (b) of models 3 and 4 are calculated differently, namely, by formula (15), the explicit form of
which is given at the end of Section 5. Then, the gravitational energies $W_3$  and $\widetilde W_3$   according to models 3
and 4 need to be calculated. First, we calculate them
according to variants (a) and (b) of model 4 by the formula [11]:
$$
 a) \quad    W_3 = -\, U_G(\nu) M\,(1), \quad 
 b) \quad \widetilde W_3 = -\, U_G(\nu)\left[M\,(1) + \widetilde M_G(1)\right]
$$
respectively. The total masses  $M\,(1)$ and $\widetilde M_G(1)$  of both
BM and DM in the LP EG have been determined in Section 3, and the potential $U_G(\nu)$ of an inner point of
the homeoid cavity is equal to [11]
$$
U_G(\nu)  = \pi G \widetilde a \widetilde c^2 J_0 \int \limits_{\nu^2}^1 \rho_G(m) dm^2
 = \frac{2 \pi G K \widetilde a \widetilde c^2J_0}{\xi (1 + \xi)}\, \frac{1 - \nu}{1 + \nu \xi} = \mbox{const}. 
$$
Here, an internal point of the homeoid interior means
any passively gravitating body (e.g., a star) located in
the LP EG. Next, setting  $\nu = \mu$ in the expression for
the potential $U_G(\nu)$, the sought-for energies $W_3$ and $\widetilde W_3$  
can be represented as
$$
 a) \quad   W_3 = -\,W_0\rho_0g_1(1), \quad  b) \quad  \widetilde W_3 = -\,W_0\left[\rho_0g_1(1) + K\,g_2(1)\right],
  \eqno (25) $$
where the function $g_1(m)$ is defined by equality (12). In
addition,
$$
  W_0 =  \frac{8 \pi^2G K \widetilde a \widetilde c^2a c^2  J_0}{\xi\,(1 + \xi)}
 \,\frac{\mu\,(1 - \mu)}{1 + \mu\,\xi}, \quad 
  g_2(m) = \frac{1}{\xi^3}\,\left[\ln (1 + \xi\,m) - \frac{\xi\,m}{1 + \xi\,m}\right]
\eqno (26) $$
 Now, let us calculate the mutual potential energies  $W_3$ and  $\widetilde W_3$
 according to variants (a) and (b) of model 3. By
analogy with formula (25), we obtain
$$
 a) \quad    W_3 = -\, U_S M\,(1), \quad b) \quad \widetilde W_3 = -\, U_S\left[M\,(1) + \widetilde M_G(1)\right], 
$$
where  $U_S$  is the potential at the inner point of the cav-
ity of the spherical layer, equal to [20]
$$
   U_S =   \di 4 \pi G  \int \limits_a^{r_s} u \, \rho_S(u) du  =
    2 \pi G K r_s^2\, \frac{r_s - a}{r_s + a} = \mbox{const},
 $$
Then 
$$
 a) \  W_3 = -\,W_0\rho_0g_1(1), \quad  b) \  \widetilde W_3 = -\,W_0\left[\rho_0g_1(1) + K\,g_2(1)\right],
 \ W_0 = 8 \pi^2 G K r_s^2\,ac^2 \frac{r_s - a}{r_s + a}
  \eqno (27) $$
 The function  $g_2(1)$ is defined above.
 
In conclusion, let us present a list of the total grav-
itational, $W$ , and kinetic, $T$, energies calculated
according to models 3, 4, and 5 in the following order.

Model 3: (a) $(a) \ W = W_L + W_S + W_3$ and $T = T_L + T_S$;  $(b) \ \widetilde W = \widetilde W_L + W_S + \widetilde W_3$,   and $\widetilde T = \widetilde T_L + T_S$. The energies
   $W_L$ and $T_L$  are determined by equality (14); $W_L$ and $T_L$,
by formulas (23) and (24), respectively; $\widetilde W_L$ and $\widetilde T_L$,  by
formula (15); and the energies $W_3$ and $\widetilde W_3$, by equality(27).
 
Model 4: (a) $(a) \ W = W_L + W_G + W_3$ and $T = T_L + T_G$;  $(b) \ \widetilde W = \widetilde W_L + W_G + \widetilde W_3$,   and $\widetilde T = \widetilde T_L + T_G$. The energies
   $W_G$ and $T_G$  are determined by formula (22), and  the energies $W_3$ and $\widetilde W_3$, by equality(25).

Model 5: $(2) \ W = \widetilde W_L$ and $T = \widetilde T_L$. Here, the energies 
 $\widetilde W_L$ and $\widetilde T_L$   are determined by formula (15). In this
case, to obtain the necessary expressions corresponding to variant (1) of model 5, it is sufficient to replace
the semiaxes $a$ and $b = c$ in expression (15) with  $\widetilde a$ and  $\widetilde b = \widetilde c$, respectively.

\vskip 4.0ex

6. VELOCITY DISPERSION OF A LAYERED 
INHOMOGENEOUS ELLIPTICAL GALAXY

\vskip 2.0ex 

Let us consider the ratio of the total rotational
energy  $T_{rot}$  of a layered inhomogeneous ellipsoid with
semiaxes $a > b > c$ and density  $\rho\,(m)$ to the absolute
value of its total gravitational energy $W$. According to
formula (7),  $W = W\,(1) = -\,\pi G a b c J_0\Psi\,(1)$; for $T_{rot} = T_1$,
we find the value of $t_z$  [11]:
$$
  t_z = \frac{T_{rot}}{|W|} = \frac{1}{2}\,\left(1 - \frac{3c^2K_0}{J_0} \right) =   \frac{J_1}{2J_0},
\eqno (28)$$
which exactly coincides with the cognominal relation
for classical homogeneous equilibrium figures
(Maclaurin spheroids: $a = b \ge c$, or Jacobi ellipsoids: $a \ge b \ge c$) and is a function of only the eccentricities
of the layer [11].

It is important to note that the ratio $t_z$ plays a key
role in establishing equilibrium and stability of an
axisymmetric dynamical system (e.g., homogeneous
Maclaurin spheroids). A criterion for the stability of
such a system, according to the Peebles–Ostriker
hypothesis [21, 22], is the fulfillment of the inequality
$$
    t_z  <  t_{crit} \approx 0.14 \pm 0.03.    
$$
Next, the ratio $t_z$  can be related with the rotational
velocity $v_{rot}(R)$ and velocity dispersion $\sigma_s(R)$ observed
in elliptical galaxies at a distance  [11]:
$$
   \frac{v_{rot}(R)}{\sigma_s(R)} = \di \sqrt{\frac{t_z}{0.5 - t_z}}
$$
The ratio $v_{rot}(R)/\sigma_s(R)$ calculated by this formula
should be compared with observations, which allows
one to draw some conclusions about the dynamic state
of this galaxy. In addition, calculating the linear rotation velocity 
$v_{rot}(R) = \Omega \, R$, we find the velocity spatial
dispersion  $\sigma_s(R)$ as a function on the distance $R$ from
the center of the galaxy [11]:
$$
  \sigma_s(R) =  \Omega  R \cdot  \sqrt{\di \frac{0.5 - t_z}{t_z} }, \quad    \Omega  = \sqrt{\frac{G M}{a^3}}
\eqno (29) $$
where $\Omega$  is the angular rotation velocity of the galaxy.
Since the centripetal and gravitational forces acting on
a point located at a distance $R$ from the center balance
each other, we obtain:
$$
   \frac{V^2}{R} = \frac{GM}{R^2}, \quad  V_{max} =  \sqrt{\frac{G M}{a}} = \Omega \, a
\eqno (30)$$
Let us present other methods for determining the
velocity spatial dispersion. Knowing the density distri-
bution law $\rho\,(r)$, one can determine the velocity spatial
dispersion $\sigma_s(R)$  at a distance $R$  from the center of the
galaxy [23, 24]:
$$
   \sigma_s^2(R) = \frac{G}{\rho\,(R)}\,\int \limits_R^{\infty} \, \frac{\rho\,(r)\, M\,(r)}{r^2}\, dr,
\eqno (31) $$
where the mass $M\,(r)$ of the intermediate ball and the
rotation velocity $ V_c(r)$ are determined by the expressions
$$
  M\,(r) = 4 \pi \,\int \limits_0^r u^2 \rho\,(u) du,  \quad V_c(r) = \di \sqrt{\frac{G M\,(r)}{r}}
\eqno (32) $$
respectively.

Another expression for determining the velocity
spatial dispersion $\sigma_s(R)$ at a distance $R$ from the center
of a spherically symmetric galaxy is given in [25]:
$$
   \sigma_s^2(R) = \frac{2G}{I\,(R)}\,\int \limits_R^{\infty}
    \, \frac{\di \sqrt{r^2 - R^2}}{r^2}\,\rho\,(r)\, M\,(r)\, dr,\quad 
    I\,(R) =  \frac{2}{\Upsilon}\, \int \limits_R^{\infty}
      \, \frac{\rho\,(r)\, r\, dr}{\di \sqrt{r^2 - R^2}},
\eqno (33) $$
where $I\,(R)$ is the surface brightness [26] and $\Upsilon$ is the
mass–luminosity ratio.

It is important to note that, for an EG with a
homothetic (astrophysical) density distribution (1),
the velocity spatial dispersion $\sigma_s(R)$ cannot be determined by formulas (31) and (33). To apply these 
formulas, we set
$$
   m = \frac{r}{q}, \quad m_{eff} = \frac{R_{eff}}{q}, \quad q = \sqrt[3]{ac^2}, \quad 
   w = \frac{r}{q}\,\sqrt{\beta} = m \sqrt{\beta},    \quad W = \frac{R\, \sqrt{\beta}}{q}.
 \eqno (34) $$
 Then, expression (31), by virtue of expressions (1)
and (15) for  $\rho\,(m)$ and $M\,(m)$   takes the form
$$
 \sigma_s^2(R) = \frac{4\pi G \rho_0 \widetilde R^3}{q\,\beta} H\,(R),\quad 
 \widetilde R =  \sqrt{q^2 + R^2\beta}
 \eqno (35) $$
where
$$
  H\,(R) =  \frac{4Q^4}{Q^4 - 1}\,\ln Q -  2 \ln (Q^2 + 1) +  \frac{2Q^2}{(Q^2+1)^2}, 
\quad  Q =  \frac{R\,\sqrt{\beta} + \widetilde R}{q}
$$
In accordance with substitution (34), from (33), we
find $I\,(R)$ and $\sigma_s^2(R)$
$$
    I\,(R) = \di \frac{2\rho_0 q^3}{\di \Upsilon \sqrt{\beta}\, \widetilde R^2}, \quad  
  \sigma_s^2(R) =  \frac{4\pi G \rho_0 \widetilde R^2\Upsilon}{\beta}\,
   \left[ K_1(R) + K_2(R) \right],
\eqno (36) $$
where
 $$
   K_1(R) =  \int \limits_W^{\infty}  \, \frac{\di \sqrt{w^2 - W^2}}{w^2 (1 + w^2)^{3/2}}\, \ln \,\left(w + \sqrt{1 + w^2}\right) \, dw, \quad 
   K_2(R) =  -\,\frac{\pi\,\left(\widetilde R - R\,\sqrt{\beta}\right)^2}{\di 4 \,\widetilde R^2}
$$

Here, the integral $K_1(R)$ cannot be calculated explicitly, but can be calculated numerically.

Formulas (35) and (36) of the velocity dispersion
correspond to variant (a) of models 3 and 4. In this
case, the LP SP does not contain DM; therefore, the
total velocity dispersion is $\sigma\,(R) \equiv \sigma_s(R)$.

Now let us determine the velocity dispersion $\sigma\,(R)$
according to variant (b) of models 3 and 4 and according to variants (1) and (2) of model 5. In this case, the
LP SP contains DM and the velocity dispersion can be
represented as the sum $\sigma^2(R) = \sigma_s^2(R) + \sigma_d^2(R)$, where
 $\sigma_s^2(R)$ is the BM component of the velocity dispersion
and $\sigma_d^2(R)$ is the DM component. In this case, it suffices to calculate $\sigma_d^2(R)$, since 
 $\sigma_s^2(R)$  is defined by formulas (35) and (36). As for a formula similar to (29), it
suffices to calculate the ratio $t_z$. According to variant (2)
of model 5, this ratio will be equal to $t_z = \widetilde T_L/|\widetilde W_L|$.
Here, 
$\widetilde W_L$ and $\widetilde T_L$  are determined by equality (15). This
ratio is easily determined according to variant (1) of
model 5 with the help of an appropriate substitution,
which was mentioned above.

Let us proceed to calculating $\sigma_d^2(R)$  first by formula(31)
$$
   \sigma_d^2(R) =  \frac{G}{\widetilde q\,\rho_G(\bar \mu)}\,\int \limits_{\bar \mu}^{\infty} \,
    \frac{\rho_G(\mu)\, M_G(\mu)}{\mu^2}\, d\mu,  \quad  \left( r = \widetilde q\, \mu, \quad 
    \bar \mu = \frac{R}{\widetilde q}, 
   \quad \widetilde q = \sqrt[3]{\widetilde a \widetilde c^2}, \quad \bar \mu \xi = \frac{R}{r_s}\right)
$$ 
Using expression (4) for the density $\rho_G(\mu)$ and mass $M_G(\mu)$  given in Section 4, we find
$$
   \sigma_d^2(R) = B_0\left[B_1(R)\ln \frac{R}{r_s} +   
   B_2(R)\ln \frac{r_s + R}{r_s} - 6B_1(R) \mbox{dilog}\,\left(\frac{r_s + R}{r_s}\right) + B_3(R)\right],
\eqno (37) $$
where 
$$
   B_1(R) =  -\,\frac{R}{r_s^4}(r_s + R)^2,\quad 
  B_2(R) = \frac{r_s + R}{r_s^4R}(r_s^3 - 3R r_s^2 + R^2r_s + 7R^3), 
$$
$$
  B_3(R) = \frac{1}{r_s^4}\left[\pi^2 R(r_s + R)^2 - r_s^3 - 9R r_s^2 - 7R^2r_s \right], \quad 
  B_0 = \frac{2\pi G K r_s \widetilde q^2}{\xi^2},
$$
and
$$
 \mbox{dilog}\,(s) = \int \limits_1^s \frac{\ln t}{1 - t}\,dt, \quad 
 \mbox{dilog}\,(1) = 0, \quad \mbox{dilog}\,(0) = \frac{\pi^2}{6}, \quad 
 \mbox{dilog}\,\left(\frac{1}{2}\right) = \frac{\pi^2}{12} - \frac{\ln^2 2}{2}
$$
Next, using the second formula (33), we determine the
surface brightness $I\,(R)$:
$$
 I\,(R) = \frac{K \widetilde q r_s^2}{\xi\,\Upsilon\,\bar R^3}\,\left(-\, \bar R + 
 r_s \ln \frac{r_s + \bar R}{r_s - \bar R}\right), \quad \bar R = \sqrt{r_s^2 - R^2}
$$
Then, in the first formula (33), we consider the
expressions for $\rho_G(\mu)$ and masses $M_G(\mu)$. Then we
determine the DM component of the velocity dispersion $\sigma_d^2(R)$:
$$
   \sigma_d^2(R) =  \frac{2 G}{I\,(R)}\,\int \limits_{\bar \mu}^{\infty} \,
    \frac{\di \sqrt{\mu^2 - \bar \mu^2}}{\mu^2}\,\rho_G(\mu)\, M_G(\mu) d\mu = 
\psi_0(R)\,\left[\psi_1(R) - \xi \, \psi_2(R)\right],
\eqno (38) $$
where 
$$
 \psi_0(R) = \frac{8 \pi G K r_s \Upsilon \, \bar R^3}{\widetilde q\,\left(-\,\bar R + 
 r_s \ln \frac{\di r_s + \bar R}{\di r_s - \bar R}\right)}, \quad
 \psi_1(R) = \int \limits_{\bar \mu}^{\infty} \,
    \frac{\di \sqrt{\mu^2 - \bar \mu^2}}{\mu^3}\,\frac{\ln (1 + \xi \mu)}{(1 + \xi \mu)^2} d\mu,   
$$
$$
 \psi_2(R) = \frac{1}{4\bar R^3r_s}\,\left[(2r_s^4 - 9 r_s^2 R^2 + 6R^4)\,
  \ln \frac{r_s + \bar R}{r_s - \bar R} - 2\bar R\,(5r_s^3 - 6 R^2r_s - 3\pi R \bar R^2) \right]
\eqno (39) $$
The integral $\psi_1(R)$ cannot be calculated analytically
and is calculated numerically.
The values of the velocity spatial dispersion$\sigma\,(R)$  at
a distance of the effective radius $R_{eff}$  for ten EGs [27]
are given in section Examples.

\vskip 4ex

7. NEW DETERMINATION OF THE DENSITY VALUES IN THE CENTER OF EG, ITS SCALE 
RADIUS, AND PARAMETER $\beta_{eff}$

\vskip 2ex

Assuming the value of the spatial velocity dispersion $\sigma_{eff} $ at a distance of the effective radius 
$R_{eff} $ to be known (see Section 6), we can determine the density $\rho_0$
 at the center of the EG. Using the astrophysical law
of density distribution $\rho\,(r)$ for the EG, according to
formula (35), we obtain
$$
 \rho_0 = \frac{\sigma_s^2(R_{eff}) \beta_{eff}\,q}{4 \pi G \sqrt{(q^2 + R^2_{eff} \beta_{eff})^3}\,H\,(R_{eff})}, \quad \left(\rho_{crit} = \frac{3 H^2}{8 \pi G} \approx 1.37 \cdot 10 ^{-7} \right)
\eqno (40) $$
In the parentheses, the critical density value in the
units  $M_{\odot}/pc^3$ is given.

If the velocity spatial dispersion  $\sigma_s(R) $ is determined by equality (36), then $\rho_0$ is expressed as follows:
$$
  \rho_0 = \frac{\sigma_s^2(R_{eff}) \beta_{eff}}
  {4 \pi G (q^2 + R^2_{eff} \beta_{eff})\,\Upsilon\,\left[K_1(R_{eff}) + K_2(R_{eff})\right]}
  \eqno (41) $$
The values of $\sigma_s(R_{eff})$, as well as the density values $\rho_0$
calculated from them at the center of the galaxy, are
given in Section 8.

The value $\beta_{eff}$  of the parameter  $\beta$ in equalities (40)
and (41), corresponding to the effective radius, can be
determined, e.g., from the Hubble law:
$$
   I_{eff} = \frac{I_0}{1 +  \beta_{eff} m_{eff}^2}, \quad  \beta_{eff} = \frac{1}{m_{eff}^2}\,
   \left(\frac{I_0}{I_{eff}} - 1\right), \quad  m_{eff} = \frac{R_{eff}}{\di \sqrt[3]{a c^2}},
\eqno (42)  $$ 
in which the central and effective surface brightness $I_0$ 
and  $I_{eff}$ are assumed to be known. Another way to
determine these parameters is related to the following
formula:
$$
  L\,(m) = 4 a q E\,(e)\int \limits_0^m m\,I\,(m)\,dm = \frac{2qa E\,(e) I_0}{\beta}\,\ln (1 + \beta\,m^2), 
  \quad  e = \di \frac{\sqrt{a^2 - c^2}}{a},
$$
in which the expression (1) for the surface brightness
 $I\,(m)$ is taken into account. Here, $L\,(m)$ is the total surface brightness of the intermediate ellipse with 
 semiaxes  $ma$ and $mc$   and  $E\,(e)$  is the complete elliptic integral of the second kind. Hence, it is easy to determine
the total, $L_T = L\,(m = 1)$, and effective, $L_{eff} = L\,(m_{eff})$, luminosity:
$$
   L_T = 2qa E\,(e) I_0\, \frac{\ln(1 + \beta_T)}{\beta_T}, \quad 
  L_{eff} = 2qa E\,(e) I_0\, \frac{\ln(1 + \beta_{eff}m_{eff}^2)}{\beta_{eff}},
$$ 
where $\beta_T$ is the value of the parameter$\beta$  corresponding
to $m = 1$. In addition, by definition, $L_{eff} = L_T/2$, i.e.,
$$
    \frac{\ln(1 + \beta_{eff}m_{eff}^2)}{\beta_{eff}} = \frac{\ln(1 + \beta_T)}{2\beta_T}
   = \frac{L_T}{4qa E\,(e) I_0}
$$ 
Hence, assuming  $I_0$ and $L_T$  to be known, we determine the parameters  $\beta_T$ and $\beta_{eff}$. In addition, the
parameter $\beta_T$ can be calculated differently. For example, if the stellar mass $M^*$ of 
the galaxy, its density  $\rho_0$ at the center, and the semiaxes $a, b = c$ are known,
then, from formula (15), for the mass $M\,(m)$, we determine the value of the parameter $\beta_T$ as a solution of the
equation
$$
    \frac{g_1(\beta_T)}{\di \beta_T \sqrt{\beta_T}} = \frac{M^*}{4 \pi \rho_0a c^2}, \quad M^* \equiv M\,(m = 1),
$$
where the function $g_1(\beta_T) \equiv g_1(m = 1)$ is
defined by equality (12).

The scale radius $r_s$ of the EG can be determined
from the condition
$$
  M_h = M^* +  M_G  +  M_S(r_s) = M_{bm}  +  M_S(r_s) = M_{bm}  +  M_{dm}
$$ 
Here, the stellar mass $M^*$, the mass of the gas $M_G$
(equal to the sum of the BM $M_{bm}$), and the mass of the
galactic halo $M_h$ are assumed to be known. In this
case, the total mass  $M_{dm} = M_S(r_s)$ of an inhomogeneous layer (model 3) consisting of DM is determined
by equality (25) at $r = r_s$. In expression (25), the normalization coefficient $K$ is replaced by its approximate value
$$
K  \approx \left(\frac{r_s}{10}\right)^{-\,2/3},
\eqno (43) $$
where $r_s$ is expressed in pc [13]. Then, assuming the
semi-major axis $a$ of the EG to be known, we represent the mass $M_{dm} = M_S(r_s)$ as a function of only the
scale radius and substitute it into the expression for $M_h$. This gives us an equation for determining $r_s$:
$$
 M_h - M_{bm} =  B\,(r_s)\,r_s^{7/3}, \quad 
 B\,(r_s) = 4 \pi \,\sqrt[3]{100}\,  \left[\ln \frac{2r_s}{a + r_s} -  \frac{r_s - a}{2(a + r_s)} \right]
\eqno (44) $$
The value of $r_s$ determined from Eq. (44) corresponds
to model 3.

In model 4, the value of $r_s$ is found from the expression for $M_h$ when $M_{dm} = M_G(\mu)$:
$$
 M_h - M_{bm} =  A\,(r_s)\,r_s^{7/3},  \quad A\,(r_s) = 4 \pi \,\sqrt[3]{100}\, 
 \left[\ln \frac{1 + \xi}{1 + \xi\,\mu} -  \frac{\xi\,(1 - \mu)}{(1 + \xi)(1 + \xi\,\mu)} \right]
\eqno(45) $$
Here 
$$
 \xi = \di \frac{\sqrt[3]{\widetilde a \widetilde c^2}}{r_s} = \di \frac{\sqrt[3]{ a c^2}}{a}, \quad \mu = \frac{a}{r_s}
$$
Now, assuming to be known $r_s$ (in pc), we determine
the normalization coefficient $K$, expressed in solar
masses per cubic parsec, using formula (43). The values of the scale radii $r_s$ thus found and the coefficient
 $K$ according to formula (43) in models 3 and 4 for ten
EGs are given in section Examples.

Thus, the formulas given in this section make it
possible to determine the values of such key parameters of an EG as the density  at the center of the galaxy and its scale radius $r_s$. The values of $\rho_0$ and $r_s$ of
specific EGs presented in the database or in works by
other authors are not given. In [28], the scale radius $\rho_s$
of am EG is understood as the radius of its disk: the
radial scale. In this case, the angular values of the
minor, $c$, and major, $a$, axes are determined from
observations. To determine these parameters, accord-
ing to the above formulas, it is necessary to know the
stellar (or baryon) mass and halo mass of the galaxy, its
effective radius, and the velocity dispersion calculated
at the distance of the effective radius of the EG. These
data are available in the database and publications.

\vskip 4 ex

8. EXAMPLES
\vskip 2.0ex

We took ten elliptical galaxies, for example, with
arameters necessary for the calculations given in
Tables 1–4.

The quantities $D_{25}$ and $R_{25}$, which are measured in
he photometric $B$  band up to a distance with a limit-
ng brightness of 25 magnitudes per square arcsecond,
s well as the heliocentric distance $D$, are taken from
he HyperLeda database (leda.union-lyonl.fr).

 Using the values of  $D_{25}$ and $R_{25}$
 given in [7], we determine the apparent values of
he major and minor semiaxes $a$ and $c$ of the EG,
which is regarded as an elongated spheroid with semixes $a > b = c$. For such galaxies and galaxies that
ave the shape of a triaxial ellipsoid, the rotation axis
oes not coincide with the apparent minor axis and
he isophotes will be misaligned. The variant of an EG
n the form of an oblate spheroid—a more convenient
r simpler variant, for which the visible axis coincides
with the axis of rotation and the isophote alignment is
ot violated—is not considered.

Table 1 presents the values of the heliocentric distance $D$ (in Mpc), BM $M_{bm}$, DM mass $M_{dm}$, and halo
mass $M_h$, scale radius $r_s$ (in kpc), as well as the normalization coefficient $K$ 
(in $10^{-3}M_{\odot}/pc^3$), determined using models 3 and 4. The masses $M_{bm}$, $M_{dm}$,
and $M_h$ are expressed in $10^{12}$ solar masses.
 
Table 2 presents the values of the EG semi-major,
$a$, and semi-minor, $c$ axes (in kpc), parameters $\beta_T$,
$m_{eff}$, and $\beta_{eff}$, as well as the effective radius $R_{eff}$
(in kpc) from [29, 30]. The parameters $\beta_T$ and $\beta_{eff}$ correspond to $m = 1$  and $m = m_{eff}$, and the latter corresponds to $R_{eff}$.

The stellar masses $M^*$, BMs $M_{bm}$, and halo masses
 $M_h$ of galaxies are given in [29–32]. For $M_h$, it is possible to use its 
 approximate dependence on the stellar
mass of the galaxy, $M^*$, $  M_h = (56 + \Delta M) \cdot M^*$, where
$\Delta M$ varies from $\Delta M = -10$ to 16 [33]. In this case, the
BMs $M_{bm}$ of EGs can be determined from the estimates of the masses of neutral hydrogen $M_{HI}$ and
molecular, $M_{H_2}$, and central molecular,  $M_{HIc}$, gases
given in [33, 34]. In addition, in [35], an approximate
formula for determining the BM of an EG is given in
the form $ M_{bm}/M_h \approx \Omega_b/\Omega_m \approx 0.17$. 

Table 3 presents the values of the central velocity
dispersion $\sigma_0$ and the spatial velocity dispersion 
$\sigma\,(R_{eff})$ (in km/s) at the distance of the effective radius,
as well as the density  at the center of the EG. Column 1 corresponds to the value of $\sigma_s(R_{eff})$ given in
[29], and column 2, to its value calculated by formula
(29). Columns 3 and 4 in the first row give the values
of  $\sigma_s(R_{eff})$ calculated by formulas (35) and (36), i.e.,
without considering the DM component. This corresponds to variant (a): the LP EG does not contain
DM. The second row of these columns gives its values
with allowance for the DM components: 
$\sigma\,(R_{eff}) = \sqrt{\sigma_s^2(R_{eff}) + \sigma_d^2(R_{eff})}$, where $\sigma_d^2(R_{eff})$ is calculated using
formulas (37) and (38). Columns 5 and 6 correspond
to the values of $\rho_0$, and column 7, to the average $\rho_0$ 
expressed in  $M_{\odot}/pc^3$.

The discrepancies in the values of $\sigma_s(R_{eff})$ are due
to the fact that, in this work, the density is distributed
in accordance with astrophysical law (6).

Table 4 presents the values of the total gravitational
and rotational kinetic energies (in Joules) of ten elliptical galaxies calculated using models 3, 4, and 5. For
model 5, the values of these quantities in variants (1)
and (2) are given, and, for models 3 and 4, in two variants: (a) the main part of DM is contained outside the
luminous part of the EG and (b) the fractions of dark
and baryonic matter in the central regions of the galaxy are comparable.

\newpage 

 {\bf Table 1.}  Values of the heliocentric distance $D$ and baryon mass $M_{bm}$ of EGs 

 \vskip 3.0ex

\begin{tabular}{|c|c|c|c|c|c|c|c|}
\hline 
\multicolumn{1}{|c|}{} &   \multicolumn{1}{|c|}{}  &   \multicolumn{1}{|c|}{} &   \multicolumn{1}{|c|}{} &   \multicolumn{1}{|c|}{}   & \multicolumn{1}{|c|}{}  &  \multicolumn{1}{|c|}{}    &  \multicolumn{1}{|c|}{} \\
  \multicolumn{1}{|c|}{NGC}  & \multicolumn{1}{|c|}{$D, Mpc$}  & \multicolumn{1}{|c|}{$M_{bm}$}
  & \multicolumn{1}{|c|}{Models} & \multicolumn{1}{|c|}{$M_{dm}$} & \multicolumn{1}{|c|}{$M_h$} 
  & \multicolumn{1}{|c|}{$r_s$}  & \multicolumn{1}{|c|}{$K$}  \\

  \multicolumn{1}{|c|}{} &   \multicolumn{1}{|c|}{}  &   \multicolumn{1}{|c|}{} &   \multicolumn{1}{|c|}{} &   \multicolumn{1}{|c|}{}   & \multicolumn{1}{|c|}{}  &  \multicolumn{1}{|c|}{}    &  \multicolumn{1}{|c|}{} \\
     \hline   
       
   \multicolumn{1}{|c|}{4365 \ E3} & \multicolumn{1}{|c|}{23.3}   &   \multicolumn{1}{|c|}{0.6842} & 
\multicolumn{1}{|c|}{3} &   \multicolumn{1}{|c|}{14.5814} &   \multicolumn{1}{|c|}{15.2656}  & 
 \multicolumn{1}{|c|}{157.80} &  \multicolumn{1}{|c|}{1.590} \\ 

\multicolumn{1}{|c|}{} &   \multicolumn{1}{|c|}{} &  \multicolumn{1}{|c|}{}  &    
\multicolumn{1}{|c|}{4} & \multicolumn{1}{|c|}{14.5694}   & \multicolumn{1}{|c|}{15.2536}  &
\multicolumn{1}{|c|}{176.00} &  \multicolumn{1}{|c|}{1.478} \\
\hline   

 \multicolumn{1}{|c|}{4374 \ E1} & \multicolumn{1}{|c|}{18.5}   &   \multicolumn{1}{|c|}{0.7258} & 
 \multicolumn{1}{|c|}{3} & \multicolumn{1}{|c|}{15.2556} &   \multicolumn{1}{|c|}{15.9814} &  
  \multicolumn{1}{|c|}{160.60}  &  \multicolumn{1}{|c|}{1.571} \\ 
 
\multicolumn{1}{|c|}{(M 84)} &   \multicolumn{1}{|c|}{} &  \multicolumn{1}{|c|}{}  &    \multicolumn{1}{|c|}{4} & \multicolumn{1}{|c|}{15.2597}   & \multicolumn{1}{|c|}{15.9855}  
& \multicolumn{1}{|c|}{168.80} &  \multicolumn{1}{|c|}{1.520} \\
\hline   

 \multicolumn{1}{|c|}{4406 \ E3} & \multicolumn{1}{|c|}{16.8}   &   \multicolumn{1}{|c|}{0.8268}  
&  \multicolumn{1}{|c|}{3} &  \multicolumn{1}{|c|}{12.6832} &   \multicolumn{1}{|c|}{13.5100} & 
  \multicolumn{1}{|c|}{148.30}  &  \multicolumn{1}{|c|}{1.657}  \\ 

\multicolumn{1}{|c|}{(M 86)} &   \multicolumn{1}{|c|}{} &  \multicolumn{1}{|c|}{}   
&  \multicolumn{1}{|c|}{4} & \multicolumn{1}{|c|}{12.6746} &   \multicolumn{1}{|c|}{13.5014} & 
  \multicolumn{1}{|c|}{165.50}  &  \multicolumn{1}{|c|}{1.540}  \\ 

\hline   

 \multicolumn{1}{|c|}{4472 \ E2} & \multicolumn{1}{|c|}{17.1}   &   \multicolumn{1}{|c|}{1.1330} & 
\multicolumn{1}{|c|}{3} &   \multicolumn{1}{|c|}{21.0525} &   \multicolumn{1}{|c|}{22.1856}  &  
\multicolumn{1}{|c|}{184.20} &  \multicolumn{1}{|c|}{1.434} \\ 
\multicolumn{1}{|c|}{(M 49)} &   \multicolumn{1}{|c|}{} &  \multicolumn{1}{|c|}{}  &   
 \multicolumn{1}{|c|}{4} & \multicolumn{1}{|c|}{21.0657}   & \multicolumn{1}{|c|}{22.1988}  
 & \multicolumn{1}{|c|}{197.0} &  \multicolumn{1}{|c|}{1.371} \\
\hline   

 \multicolumn{1}{|c|}{4494 \ E2} & \multicolumn{1}{|c|}{16.6}   &   \multicolumn{1}{|c|}{0.1734} & 
\multicolumn{1}{|c|}{3} &   \multicolumn{1}{|c|}{4.4278} &   \multicolumn{1}{|c|}{4.6012}  &  
\multicolumn{1}{|c|}{94.470} &  \multicolumn{1}{|c|}{2.238} \\ 
\multicolumn{1}{|c|}{} &   \multicolumn{1}{|c|}{} &  \multicolumn{1}{|c|}{}  &  
  \multicolumn{1}{|c|}{4} & \multicolumn{1}{|c|}{4.4240}   & \multicolumn{1}{|c|}{4.5974}  
  & \multicolumn{1}{|c|}{102.70} &  \multicolumn{1}{|c|}{2.117} \\
\hline   

 \multicolumn{1}{|c|}{4621 \ E4} & \multicolumn{1}{|c|}{14.9}   &   \multicolumn{1}{|c|}{0.2586} & 
\multicolumn{1}{|c|}{3} &   \multicolumn{1}{|c|}{5.9330} &   \multicolumn{1}{|c|}{6.1916}  &  
\multicolumn{1}{|c|}{106.70} &  \multicolumn{1}{|c|}{2.063} \\ 
\multicolumn{1}{|c|}{} &   \multicolumn{1}{|c|}{} &  \multicolumn{1}{|c|}{}  & 
   \multicolumn{1}{|c|}{4} & \multicolumn{1}{|c|}{5.9285}   & \multicolumn{1}{|c|}{6.1871}
     & \multicolumn{1}{|c|}{124.60} &  \multicolumn{1}{|c|}{1.861} \\
\hline   

 \multicolumn{1}{|c|}{4636 \ E2} & \multicolumn{1}{|c|}{14.3}   &   \multicolumn{1}{|c|}{0.4680} & 
\multicolumn{1}{|c|}{3} &   \multicolumn{1}{|c|}{11.5722} &   \multicolumn{1}{|c|}{12.0402}  &  
\multicolumn{1}{|c|}{141.70} &  \multicolumn{1}{|c|}{1.708} \\ 
\multicolumn{1}{|c|}{} &   \multicolumn{1}{|c|}{} &  \multicolumn{1}{|c|}{}  &   
 \multicolumn{1}{|c|}{4} & \multicolumn{1}{|c|}{11.5560}   & \multicolumn{1}{|c|}{12.0239}  & 
 \multicolumn{1}{|c|}{152.90} &  \multicolumn{1}{|c|}{1.623} \\
\hline   

 \multicolumn{1}{|c|}{4649 \ E2} & \multicolumn{1}{|c|}{17.3}   &   \multicolumn{1}{|c|}{0.9479} & 
\multicolumn{1}{|c|}{3} &   \multicolumn{1}{|c|}{17.9113} &   \multicolumn{1}{|c|}{18.8593}  &  
\multicolumn{1}{|c|}{171.40} &  \multicolumn{1}{|c|}{1.504} \\ 
\multicolumn{1}{|c|}{(M 60)} &   \multicolumn{1}{|c|}{} &  \multicolumn{1}{|c|}{}  &   
 \multicolumn{1}{|c|}{4} & \multicolumn{1}{|c|}{17.9158}   & \multicolumn{1}{|c|}{18.8637} 
  & \multicolumn{1}{|c|}{181.80} &  \multicolumn{1}{|c|}{1.446} \\
\hline 
  
 \multicolumn{1}{|c|}{4697 \ E4} & \multicolumn{1}{|c|}{11.4}   &   \multicolumn{1}{|c|}{0.2433} & 
\multicolumn{1}{|c|}{3} &   \multicolumn{1}{|c|}{6.3532} &   \multicolumn{1}{|c|}{6.5964}  &  
\multicolumn{1}{|c|}{109.60} &  \multicolumn{1}{|c|}{2.027} \\ 
\multicolumn{1}{|c|}{} &   \multicolumn{1}{|c|}{} &  \multicolumn{1}{|c|}{}  &   
 \multicolumn{1}{|c|}{4} & \multicolumn{1}{|c|}{6.3512} & \multicolumn{1}{|c|}{6.5944}  &
  \multicolumn{1}{|c|}{130.50} &  \multicolumn{1}{|c|}{1.804} \\
\hline   
  
 \multicolumn{1}{|c|}{7454 \ E2} & \multicolumn{1}{|c|}{23.2}   &   \multicolumn{1}{|c|}{0.0741} & 
\multicolumn{1}{|c|}{3} &   \multicolumn{1}{|c|}{1.9067} &   \multicolumn{1}{|c|}{1.9808}  &  
\multicolumn{1}{|c|}{65.60} &  \multicolumn{1}{|c|}{2.854} \\ 
\multicolumn{1}{|c|}{} &   \multicolumn{1}{|c|}{} &  \multicolumn{1}{|c|}{}  &   
 \multicolumn{1}{|c|}{4} & \multicolumn{1}{|c|}{1.9061}   & \multicolumn{1}{|c|}{1.9802}  & 
 \multicolumn{1}{|c|}{72.02} &  \multicolumn{1}{|c|}{2.681} \\
\hline 

 \end{tabular}
 
  \vskip 2.0ex
{\small  The masses of dark matter, $M_{dm}$, and halo,   $M_h$, as well as the scales radius $r_s$ (in kpc) were calculated according to models 3 and 4. The masses are expressed in $10^{12}M_{\odot}$, and the normalization coefficient  $K$, in 
 $10^{-3}M_{\odot}/pc^3$.}

\newpage 

{\bf Table 2.} Values of the semi-major, $a$, and semi-minor, $c$, axes of EGs and the effective radius 
$R_{eff}$ 

  \vskip 3.0ex

\begin{tabular}{|c|c|c|c|c|c|c|}

\hline 
  \multicolumn{1}{|c|}{} & \multicolumn{1}{|c|}{} & \multicolumn{1}{|c|}{} & \multicolumn{1}{|c|}{} & 
  \multicolumn{1}{|c|}{}  & \multicolumn{1}{|c|}{}   & \multicolumn{1}{|c|}{} \\
  
  \multicolumn{1}{|c|}{NGC}  & \multicolumn{1}{|c|}{$a, kpc$} & \multicolumn{1}{|c|}{$c, kpc$} & 
  \multicolumn{1}{|c|}{$\beta_T$} & \multicolumn{1}{|c|}{$m_{eff}$}    & \multicolumn{1}{|c|}{$\beta_{eff}$}
   & \multicolumn{1}{|c|}{$R_{eff}, kpc$}   \\

  \multicolumn{1}{|c|}{} &   \multicolumn{1}{|c|}{} &     \multicolumn{1}{|c|}{} &  \multicolumn{1}{|c|}{} 
   & \multicolumn{1}{|c|}{}    & \multicolumn{1}{|c|}{}    & \multicolumn{1}{|c|}{}   \\
 \hline 
  
 \multicolumn{1}{|c|}{4365 \ E3} & \multicolumn{1}{|c|}{20.896}   &  
  \multicolumn{1}{|c|}{15.490} & \multicolumn{1}{|c|}{1485.30} &   \multicolumn{1}{|c|}{0.284} & 
   \multicolumn{1}{|c|}{2086.13}  &  \multicolumn{1}{|c|}{6.775}  \\ 
\hline   
    
 \multicolumn{1}{|c|}{4374 \ E1} & \multicolumn{1}{|c|}{19.947}   &   \multicolumn{1}{|c|}{17.373} & 
\multicolumn{1}{|c|}{2555.69} &   \multicolumn{1}{|c|}{0.223} &   \multicolumn{1}{|c|}{3333.71}  
&  \multicolumn{1}{|c|}{ 5.492} \\ 

\multicolumn{1}{|c|}{(M 84)} &   \multicolumn{1}{|c|}{} &  \multicolumn{1}{|c|}{}  &   
 \multicolumn{1}{|c|}{} & \multicolumn{1}{|c|}{}  &  \multicolumn{1}{|c|}{} & \multicolumn{1}{|c|}{}  \\
\hline   

 \multicolumn{1}{|c|}{4406 \ E3} & \multicolumn{1}{|c|}{18.114}   &   \multicolumn{1}{|c|}{13.428} &
  \multicolumn{1}{|c|}{1032.91}  &   \multicolumn{1}{|c|}{0.341} &   \multicolumn{1}{|c|}{1547.86}  & 
  \multicolumn{1}{|c|}{10.140} \\ 
  
\multicolumn{1}{|c|}{(M 86)} &   \multicolumn{1}{|c|}{} &  \multicolumn{1}{|c|}{} 
 &    \multicolumn{1}{|c|}{} & \multicolumn{1}{|c|}{}  & \multicolumn{1}{|c|}{}  & \multicolumn{1}{|c|}{} \\
\hline   

 \multicolumn{1}{|c|}{4472 \ E2} & \multicolumn{1}{|c|}{22.166}   &   \multicolumn{1}{|c|}{18.437} & 
 \multicolumn{1}{|c|}{554.84} &   \multicolumn{1}{|c|}{0.447} &   \multicolumn{1}{|c|}{915.70}  &
  \multicolumn{1}{|c|}{8.661}  \\
   
\multicolumn{1}{|c|}{(M 49)} &   \multicolumn{1}{|c|}{} &  \multicolumn{1}{|c|}{}  &    
\multicolumn{1}{|c|}{} & \multicolumn{1}{|c|}{}  & \multicolumn{1}{|c|}{}  & \multicolumn{1}{|c|}{}  \\
\hline   

 \multicolumn{1}{|c|}{4494 \ E2} & \multicolumn{1}{|c|}{11.556}   &   \multicolumn{1}{|c|}{9.179} & 
 \multicolumn{1}{|c|}{2014.97}  &   \multicolumn{1}{|c|}{0.247} &   \multicolumn{1}{|c|}{2706.64}  & 
  \multicolumn{1}{|c|}{3.764} \\
 
\hline   

 \multicolumn{1}{|c|}{4621 \ E4} & \multicolumn{1}{|c|}{11.114}   &   \multicolumn{1}{|c|}{7.343} & 
 \multicolumn{1}{|c|}{790.19}  &   \multicolumn{1}{|c|}{0.397} &   \multicolumn{1}{|c|}{1253.04}  &  
 \multicolumn{1}{|c|}{3.219}  \\
  
\hline   

 \multicolumn{1}{|c|}{4636 \ E2} & \multicolumn{1}{|c|}{12.824}   &   \multicolumn{1}{|c|}{10.424} & 
 \multicolumn{1}{|c|}{415.36}  &   \multicolumn{1}{|c|}{0.542} &   \multicolumn{1}{|c|}{742.45}  &
   \multicolumn{1}{|c|}{6.50} \\ 
   
\hline   
   
 \multicolumn{1}{|c|}{4649 \ E2} & \multicolumn{1}{|c|}{18.228}   &   \multicolumn{1}{|c|}{15.515}
 & \multicolumn{1}{|c|}{452.54}  &   \multicolumn{1}{|c|}{0.476} &   \multicolumn{1}{|c|}{763.50}  & 
  \multicolumn{1}{|c|}{6.421} \\ 
  
\multicolumn{1}{|c|}{(M 60)} &   \multicolumn{1}{|c|}{} &  \multicolumn{1}{|c|}{}  &  
  \multicolumn{1}{|c|}{} & \multicolumn{1}{|c|}{}   & \multicolumn{1}{|c|}{} 
   & \multicolumn{1}{|c|}{} \\
\hline 
  
 \multicolumn{1}{|c|}{4697 \ E4} & \multicolumn{1}{|c|}{9.991}  &   \multicolumn{1}{|c|}{6.304} &
 \multicolumn{1}{|c|}{854.44}  &   \multicolumn{1}{|c|}{0.375} &   \multicolumn{1}{|c|}{1324.18}  & 
  \multicolumn{1}{|c|}{3.922}  \\ 
  
\hline   
  
 \multicolumn{1}{|c|}{7454 \ E2} & \multicolumn{1}{|c|}{6.889}   &   \multicolumn{1}{|c|}{5.348} 
 & \multicolumn{1}{|c|}{913.44}  &   \multicolumn{1}{|c|}{0.346} &   \multicolumn{1}{|c|}{1366.87}  & 
  \multicolumn{1}{|c|}{2.692}  \\ 
  
 \hline   
 \end{tabular}

 \vskip 2.0ex
 {\small The quantities $\beta_T$ and $\beta_{eff}$ correspond to $m = 1$ and $m = m_{eff}$.} 
 
 \newpage 

{\bf Table 3.} Values of the central, $\sigma_0$, and spatial, $\sigma\,(R_{eff})$, velocity dispersion at the distance of the effective radius and density  $\rho_0$ in the center of EGs

 \vskip 2.0ex

\begin{tabular}{|c|c|c|c|c|c|c|c|c|}
\hline 
   
   \multicolumn{1}{|c|}{NGC} &   \multicolumn{1}{|c|}{$\sigma_0, km/s$} &  
    \multicolumn{4}{c}{$\sigma\,(R_{eff}), km/s$}    &  \multicolumn{3}{|c|}{$\rho_0, M_{odot}/pc^3$ }   \\
    \cline{3-9} 
   \multicolumn{1}{|c|}{} &   \multicolumn{1}{|c|}{} &  \multicolumn{1}{|c|}{1} 
   & \multicolumn{1}{|c|}{2}   & \multicolumn{1}{|c|}{3}  & \multicolumn{1}{|c|}{4}  & \multicolumn{1}{|c|}{5} & \multicolumn{1}{|c|}{6} &   \multicolumn{1}{|c|}{7}\\  
  \hline 
  
      \multicolumn{1}{|c|}{4365 \ E3} &   \multicolumn{1}{|c|}{255.90} &     \multicolumn{1}{|c|}{221.31} & 
      \multicolumn{1}{|c|}{289.33}     & \multicolumn{1}{|c|}{221.09}  & \multicolumn{1}{|c|}{118.90}
       & \multicolumn{1}{|c|}{68.194} & \multicolumn{1}{|c|}{10.73} &  \multicolumn{1}{|c|}{39.46} \\
   
 \multicolumn{1}{|c|}{} &   \multicolumn{1}{|c|}{} &  \multicolumn{1}{|c|}{}  &    
\multicolumn{1}{|c|}{} & \multicolumn{1}{|c|}{387.66} & \multicolumn{1}{|c|}{382.24}  & 
\multicolumn{1}{|c|}{}  & \multicolumn{1}{|c|}{}  &   \multicolumn{1}{|c|}{}\\

\hline   
    
 \multicolumn{1}{|c|}{4374 \ E1} & \multicolumn{1}{|c|}{288.40}   &   \multicolumn{1}{|c|}{258.23} 
 & \multicolumn{1}{|c|}{388.14} &   \multicolumn{1}{|c|}{239.80} &   \multicolumn{1}{|c|}{128.80}
   &  \multicolumn{1}{|c|}{132.71} & \multicolumn{1}{|c|}{17.52} & \multicolumn{1}{|c|}{75.113} \\ 

\multicolumn{1}{|c|}{(M 84)} &   \multicolumn{1}{|c|}{} &  \multicolumn{1}{|c|}{}  &  
  \multicolumn{1}{|c|}{} & \multicolumn{1}{|c|}{377.34} &  \multicolumn{1}{|c|}{371.51} & \multicolumn{1}{|c|}{} & \multicolumn{1}{|c|}{}  &   \multicolumn{1}{|c|}{} \\
\hline   

 \multicolumn{1}{|c|}{4406 \ E3} & \multicolumn{1}{|c|}{216.80}   &   \multicolumn{1}{|c|}{190.55} & 
 \multicolumn{1}{|c|}{500.52} &   \multicolumn{1}{|c|}{219.23} &   \multicolumn{1}{|c|}{118.61}
   &  \multicolumn{1}{|c|}{22.58} &  \multicolumn{1}{|c|}{3.45}&   \multicolumn{1}{|c|}{13.01380} \\ 
   
\multicolumn{1}{|c|}{(M 86)} &   \multicolumn{1}{|c|}{} &  \multicolumn{1}{|c|}{}  &   
 \multicolumn{1}{|c|}{} & \multicolumn{1}{|c|}{426.45}   &   \multicolumn{1}{|c|}{380.13}
& \multicolumn{1}{|c|}{}  & \multicolumn{1}{|c|}{} &  \multicolumn{1}{|c|}{} \\
\hline   

 \multicolumn{1}{|c|}{4472 \ E2} & \multicolumn{1}{|c|}{288.30}   &   \multicolumn{1}{|c|}{250.04}
  & \multicolumn{1}{|c|}{527.89} &   \multicolumn{1}{|c|}{255.82} &   \multicolumn{1}{|c|}{137.85}  &
 \multicolumn{1}{|c|}{50.844}  &  \multicolumn{1}{|c|}{7.89}&   \multicolumn{1}{|c|}{29.366}\\ 
 
\multicolumn{1}{|c|}{(M 49)} &   \multicolumn{1}{|c|}{} &  \multicolumn{1}{|c|}{}  &  
  \multicolumn{1}{|c|}{} & \multicolumn{1}{|c|}{442.19} &   \multicolumn{1}{|c|}{419.12}
 & \multicolumn{1}{|c|}{}  & \multicolumn{1}{|c|}{} &  \multicolumn{1}{|c|}{} \\
\hline   

 \multicolumn{1}{|c|}{4494 \ E2} & \multicolumn{1}{|c|}{261.80}   &   \multicolumn{1}{|c|}{224.39} &
  \multicolumn{1}{|c|}{248.31}  &   \multicolumn{1}{|c|}{152.10} &   \multicolumn{1}{|c|}{81.82}  & 
   \multicolumn{1}{|c|}{98.34} &  \multicolumn{1}{|c|}{21.42} &   \multicolumn{1}{|c|}{59.878} \\ 
   
\multicolumn{1}{|c|}{} &   \multicolumn{1}{|c|}{} &  \multicolumn{1}{|c|}{}  & 
   \multicolumn{1}{|c|}{} & \multicolumn{1}{|c|}{266.16} &   \multicolumn{1}{|c|}{265.60}
  & \multicolumn{1}{|c|}{}  & \multicolumn{1}{|c|}{}  &  \multicolumn{1}{|c|}{}\\
\hline   

 \multicolumn{1}{|c|}{4621 \ E4} & \multicolumn{1}{|c|}{223.90}   &    \multicolumn{1}{|c|}{197.69} & 
 \multicolumn{1}{|c|}{188.99}  &   \multicolumn{1}{|c|}{195.54} &   \multicolumn{1}{|c|}{105.02}  &  
 \multicolumn{1}{|c|}{254.57} &  \multicolumn{1}{|c|}{37.97} &   \multicolumn{1}{|c|}{146.27}\\ 
 
\multicolumn{1}{|c|}{} &   \multicolumn{1}{|c|}{} &  \multicolumn{1}{|c|}{}  & 
   \multicolumn{1}{|c|}{} & \multicolumn{1}{|c|}{293.71}  &   \multicolumn{1}{|c|}{298.20} 
& \multicolumn{1}{|c|}{}  & \multicolumn{1}{|c|}{}  &  \multicolumn{1}{|c|}{}\\
\hline   

 \multicolumn{1}{|c|}{4636 \ E2} & \multicolumn{1}{|c|}{199.50}   &    \multicolumn{1}{|c|}{181.55} & 
 \multicolumn{1}{|c|}{627.01}  &   \multicolumn{1}{|c|}{195.39} &   \multicolumn{1}{|c|}{105.58}  & 
   \multicolumn{1}{|c|}{48.12} &  \multicolumn{1}{|c|}{5.18}&   \multicolumn{1}{|c|}{26.650}\\
    
\multicolumn{1}{|c|}{} &   \multicolumn{1}{|c|}{} &  \multicolumn{1}{|c|}{}  &  
  \multicolumn{1}{|c|}{} & \multicolumn{1}{|c|}{358.66}   &   \multicolumn{1}{|c|}{349.67}
& \multicolumn{1}{|c|}{}  & \multicolumn{1}{|c|}{}  &  \multicolumn{1}{|c|}{} \\
\hline   
   
 \multicolumn{1}{|c|}{4649 \ E2} & \multicolumn{1}{|c|}{314.80}  
  &   \multicolumn{1}{|c|}{267.92} & \multicolumn{1}{|c|}{519.45}
  &   \multicolumn{1}{|c|}{262.99} &   \multicolumn{1}{|c|}{141.57}  &
    \multicolumn{1}{|c|}{105.36} &  \multicolumn{1}{|c|}{13.22} &   \multicolumn{1}{|c|}{59.289} \\ 
    
\multicolumn{1}{|c|}{(M 60)} &   \multicolumn{1}{|c|}{} &  \multicolumn{1}{|c|}{}  & 
   \multicolumn{1}{|c|}{} & \multicolumn{1}{|c|}{410.56}  &   \multicolumn{1}{|c|}{394.82}
 & \multicolumn{1}{|c|}{}  & \multicolumn{1}{|c|}{} &  \multicolumn{1}{|c|}{}\\
\hline 
  
 \multicolumn{1}{|c|}{4697 \ E4} & \multicolumn{1}{|c|}{180.70}  
 &   \multicolumn{1}{|c|}{169.43} & \multicolumn{1}{|c|}{263.52} 
 &   \multicolumn{1}{|c|}{183.70} &   \multicolumn{1}{|c|}{98.93}  & 
  \multicolumn{1}{|c|}{129.73} &  \multicolumn{1}{|c|}{22.31} &   \multicolumn{1}{|c|}{76.017}\\
   
\multicolumn{1}{|c|}{} &   \multicolumn{1}{|c|}{} &  \multicolumn{1}{|c|}{}  &  
  \multicolumn{1}{|c|}{} & \multicolumn{1}{|c|}{301.03} &   \multicolumn{1}{|c|}{307.87}
  & \multicolumn{1}{|c|}{}  & \multicolumn{1}{|c|}{}  &  \multicolumn{1}{|c|}{}\\
\hline   
  
 \multicolumn{1}{|c|}{7454 \ E2} & \multicolumn{1}{|c|}{231.70}  
  &   \multicolumn{1}{|c|}{223.36}& \multicolumn{1}{|c|}{240.49} 
 &   \multicolumn{1}{|c|}{118.60} &   \multicolumn{1}{|c|}{63.90}  &  
 \multicolumn{1}{|c|}{113.05} &  \multicolumn{1}{|c|}{18.10}  &   \multicolumn{1}{|c|}{65.572}\\ 
 
\multicolumn{1}{|c|}{} &   \multicolumn{1}{|c|}{} &  \multicolumn{1}{|c|}{}  &  
  \multicolumn{1}{|c|}{} & \multicolumn{1}{|c|}{210.25}   &   \multicolumn{1}{|c|}{209.71}
& \multicolumn{1}{|c|}{}  & \multicolumn{1}{|c|}{}  &  \multicolumn{1}{|c|}{}\\
 \hline   
 \end{tabular}

\vskip 2.0ex

{\small Columns 1 and 2 correspond to the values of $\sigma\,(R_{eff}$ given in [29] and calculated according to (29). Columns 3 and 4 in the first row givethe values of $\sigma_s(R_{eff}$  taking into account only the BM component, and in the second column, the values of $\sigma\,(R_{eff}$ with allowance for
the DM components. Columns 5 and 6 correspond to the values of $\rho_0$, and column 7 to the average value of $\rho_0$.}

 \newpage
 
 {\bf Table 4.} Values of total gravitational energies $W$ and rotational kinetic energies $T$ of EGs, 
 calculated according to models 3, 4, and 5

\vskip 2.0ex

\begin{tabular}{|c|c|c|c|c|c|c|c|}
 \hline 

   \multicolumn{1}{|c|}{ } & \multicolumn{1}{|c|}{ } & \multicolumn{1}{c|}{} 
&    \multicolumn{1}{|c|}{ } & \multicolumn{1}{|c|}{ } & \multicolumn{1}{c|}{} &
   \multicolumn{1}{|c|}{ } & \multicolumn{1}{|c|}{ } \\
      
   \multicolumn{1}{|c|}{NGC} & \multicolumn{1}{|c|}{Models}  & \multicolumn{1}{c|}{$W$} & \multicolumn{1}{c|}{$T$}  &   \multicolumn{1}{|c|}{NGC} & \multicolumn{1}{|c|}{Models}  & \multicolumn{1}{c|}{$W$} & \multicolumn{1}{c|}{$T$}\\
  
   \multicolumn{1}{|c|}{} &   \multicolumn{1}{|c|}{} & \multicolumn{1}{c|}{} 
   & \multicolumn{1}{c|}{} &\multicolumn{1}{|c|}{} &   \multicolumn{1}{|c|}{} & 
   \multicolumn{1}{c|}{} & \multicolumn{1}{c|}{} \\
  \hline 
     
  \multicolumn{1}{|c|}{} &   \multicolumn{1}{|c|}{3} & \multicolumn{1}{c|}{a) \ $-\,1.0238$} 
 & \multicolumn{1}{c|}{$0.1727$} & \multicolumn{1}{|c|}{} &   \multicolumn{1}{|c|}{3} &
  \multicolumn{1}{c|}{a) \ $-\,0.2567$}  &  \multicolumn{1}{c|}{$0.0419$}\\
  
 \multicolumn{1}{|c|}{} &   \multicolumn{1}{|c|}{} & \multicolumn{1}{c|}{b) \ $-\,1.0303$} 
 & \multicolumn{1}{c|}{$0.1728$} & \multicolumn{1}{|c|}{} &   \multicolumn{1}{|c|}{} & 
 \multicolumn{1}{c|}{b) \ $-\,0.2574$}  & \multicolumn{1}{c|}{$0.0419$}\\

 \multicolumn{1}{|c|}{4365} &   \multicolumn{1}{|c|}{4} & \multicolumn{1}{c|}{a) \ $-\,9.4667$}  & 
 \multicolumn{1}{c|}{$0.3803$} &\multicolumn{1}{|c|}{4621} &   \multicolumn{1}{|c|}{4} &
  \multicolumn{1}{c|}{a) \ $-\,2.0220$}  & \multicolumn{1}{c|}{$0.1124$}\\
  
 \multicolumn{1}{|c|}{} &   \multicolumn{1}{|c|}{} & \multicolumn{1}{c|}{b) \ $-\,9.4689$} 
 & \multicolumn{1}{c|}{$0.3803$} &\multicolumn{1}{|c|}{} &   \multicolumn{1}{|c|}{}
  & \multicolumn{1}{c|}{b) \ $-\,2.0223$}  & \multicolumn{1}{c|}{$0.1125$}\\

 \multicolumn{1}{|c|}{} &   \multicolumn{1}{|c|}{5} & \multicolumn{1}{c|}{1) \ $-\,1951.72$}  & 
 \multicolumn{1}{c|}{$78.516$} &\multicolumn{1}{|c|}{} &   \multicolumn{1}{|c|}{5} &
  \multicolumn{1}{c|}{1) \ $-\,2372.6$}  & \multicolumn{1}{c|}{$132.12$}\\
  
 \multicolumn{1}{|c|}{} &   \multicolumn{1}{|c|}{} & \multicolumn{1}{c|}{2) \ $-\,0.0461$} 
 & \multicolumn{1}{c|}{$0.0018$} &\multicolumn{1}{|c|}{} &   \multicolumn{1}{|c|}{}
  & \multicolumn{1}{c|}{2) \ $-\,0.0139$}  & \multicolumn{1}{c|}{$0.0008$}\\

  \hline   

  \multicolumn{1}{|c|}{} &   \multicolumn{1}{|c|}{3} & \multicolumn{1}{c|}{a) \ $-\,1.1024$} 
 & \multicolumn{1}{c|}{$0.1843$} & \multicolumn{1}{|c|}{} &   \multicolumn{1}{|c|}{3} &
  \multicolumn{1}{c|}{a) \ $-\,0.7105$}  &  \multicolumn{1}{c|}{$0.1177$}\\
  
 \multicolumn{1}{|c|}{} &   \multicolumn{1}{|c|}{} & \multicolumn{1}{c|}{b) \ $-\,1.1089$} 
 & \multicolumn{1}{c|}{$0.1843$} & \multicolumn{1}{|c|}{} &   \multicolumn{1}{|c|}{} & 
 \multicolumn{1}{c|}{b) \ $-\,0.7122$}  & \multicolumn{1}{c|}{$0.1178$}\\

 \multicolumn{1}{|c|}{4374} &   \multicolumn{1}{|c|}{4} & \multicolumn{1}{c|}{a) \ $-\,12.644$}  & 
 \multicolumn{1}{c|}{$0.2339$} &\multicolumn{1}{|c|}{4636} &   \multicolumn{1}{|c|}{4} &
  \multicolumn{1}{c|}{a) \ $-\,7.8006$}  & \multicolumn{1}{c|}{$0.2168$}\\
  
 \multicolumn{1}{|c|}{} &   \multicolumn{1}{|c|}{} & \multicolumn{1}{c|}{b) \ $-\,1.1089$} 
 & \multicolumn{1}{c|}{$0.1843$} &\multicolumn{1}{|c|}{} &   \multicolumn{1}{|c|}{}
  & \multicolumn{1}{c|}{b) \ $-\,7.8012$}  & \multicolumn{1}{c|}{$0.2168$}\\

 \multicolumn{1}{|c|}{} &   \multicolumn{1}{|c|}{5} & \multicolumn{1}{c|}{1) \ $-\,2466.2$}  & 
 \multicolumn{1}{c|}{$45.666$} &\multicolumn{1}{|c|}{} &   \multicolumn{1}{|c|}{5} &
  \multicolumn{1}{c|}{1) \ $-\,6223.4$}  & \multicolumn{1}{c|}{$173.13$}\\
  
 \multicolumn{1}{|c|}{} &   \multicolumn{1}{|c|}{} & \multicolumn{1}{c|}{2) \ $-\,0.0568$} 
 & \multicolumn{1}{c|}{$0.0011$} &\multicolumn{1}{|c|}{} &   \multicolumn{1}{|c|}{}
  & \multicolumn{1}{c|}{2) \ $-\,0.0258$}  & \multicolumn{1}{c|}{$0.0007$}\\

  \hline    
  
  \multicolumn{1}{|c|}{} &   \multicolumn{1}{|c|}{3} & \multicolumn{1}{c|}{a) \ $-\,0.8899$} 
 & \multicolumn{1}{c|}{$0.1415$} & \multicolumn{1}{|c|}{} &   \multicolumn{1}{|c|}{3} &
  \multicolumn{1}{c|}{a) \ $-\,1.4781$}  &  \multicolumn{1}{c|}{$0.2379$}\\
  
 \multicolumn{1}{|c|}{} &   \multicolumn{1}{|c|}{} & \multicolumn{1}{c|}{b) \ $-\,0.8946$} 
 & \multicolumn{1}{c|}{$0.1416$} & \multicolumn{1}{|c|}{} &   \multicolumn{1}{|c|}{} & 
 \multicolumn{1}{c|}{b) \ $-\,1.4836$}  & \multicolumn{1}{c|}{$0.2379$}\\

 \multicolumn{1}{|c|}{4406} &   \multicolumn{1}{|c|}{4} & \multicolumn{1}{c|}{a) \ $-\,7.7024$}  & 
 \multicolumn{1}{c|}{$0.3093$} &\multicolumn{1}{|c|}{4649} &   \multicolumn{1}{|c|}{4} &
  \multicolumn{1}{c|}{a) \ $-\,16.171$}  & \multicolumn{1}{c|}{$0.3492$}\\
  
 \multicolumn{1}{|c|}{} &   \multicolumn{1}{|c|}{} & \multicolumn{1}{c|}{b) \ $-\,7.7042$} 
 & \multicolumn{1}{c|}{$0.3093$} &\multicolumn{1}{|c|}{} &   \multicolumn{1}{|c|}{}
  & \multicolumn{1}{c|}{b) \ $-\,16.173$}  & \multicolumn{1}{c|}{$0.3492$}\\

 \multicolumn{1}{|c|}{} &   \multicolumn{1}{|c|}{5} & \multicolumn{1}{c|}{1) \ $-\,3570.2$}  & 
 \multicolumn{1}{c|}{$143.63$} &\multicolumn{1}{|c|}{} &   \multicolumn{1}{|c|}{5} &
  \multicolumn{1}{c|}{1) \ $-\,897.39$}  & \multicolumn{1}{c|}{$193.99$}\\
  
 \multicolumn{1}{|c|}{} &   \multicolumn{1}{|c|}{} & \multicolumn{1}{c|}{2) \ $-\,0.0561$} 
 & \multicolumn{1}{c|}{$0.0023$} &\multicolumn{1}{|c|}{} &   \multicolumn{1}{|c|}{}
  & \multicolumn{1}{c|}{2) \ $-\,0.0909$}  & \multicolumn{1}{c|}{$0.0020$}\\

  \hline   
  
  \multicolumn{1}{|c|}{} &   \multicolumn{1}{|c|}{3} & \multicolumn{1}{c|}{a) \ $-\,1.8781$} 
 & \multicolumn{1}{c|}{$0.3079$} & \multicolumn{1}{|c|}{} &   \multicolumn{1}{|c|}{3} &
  \multicolumn{1}{c|}{a) \ $-\,0.2809$}  &  \multicolumn{1}{c|}{$0.0462$}\\
  
 \multicolumn{1}{|c|}{} &   \multicolumn{1}{|c|}{} & \multicolumn{1}{c|}{b) \ $-\,1.8881$} 
 & \multicolumn{1}{c|}{$0.3081$} & \multicolumn{1}{|c|}{} &   \multicolumn{1}{|c|}{} & 
 \multicolumn{1}{c|}{b) \ $-\,0.2815$}  & \multicolumn{1}{c|}{$0.04621$}\\

 \multicolumn{1}{|c|}{4472} &   \multicolumn{1}{|c|}{4} & \multicolumn{1}{c|}{a) \ $-\,19.874$}  & 
 \multicolumn{1}{c|}{$0.4906$} &\multicolumn{1}{|c|}{4697} &   \multicolumn{1}{|c|}{4} &
  \multicolumn{1}{c|}{a) \ $-\,2.1293$}  & \multicolumn{1}{c|}{$0.1315$}\\
  
 \multicolumn{1}{|c|}{} &   \multicolumn{1}{|c|}{} & \multicolumn{1}{c|}{b) \ $-\,19.877$} 
 & \multicolumn{1}{c|}{$0.4906$} &\multicolumn{1}{|c|}{} &   \multicolumn{1}{|c|}{}
  & \multicolumn{1}{c|}{b) \ $-\,2.1295$}  & \multicolumn{1}{c|}{$0.1315$}\\

 \multicolumn{1}{|c|}{} &   \multicolumn{1}{|c|}{5} & \multicolumn{1}{c|}{1) \ $-\,5665.061$}  & 
 \multicolumn{1}{c|}{$140.03$} &\multicolumn{1}{|c|}{} &   \multicolumn{1}{|c|}{5} &
  \multicolumn{1}{c|}{1) \ $-\,4244.45$}  & \multicolumn{1}{c|}{$262.47$}\\
  
 \multicolumn{1}{|c|}{} &   \multicolumn{1}{|c|}{} & \multicolumn{1}{c|}{2) \ $-\,0.1022$} 
 & \multicolumn{1}{c|}{$0.00253$} &\multicolumn{1}{|c|}{} &   \multicolumn{1}{|c|}{}
  & \multicolumn{1}{c|}{2) \ $-\,0.0111$}  & \multicolumn{1}{c|}{$0.0007$}\\

  \hline      

  \multicolumn{1}{|c|}{} &   \multicolumn{1}{|c|}{3} & \multicolumn{1}{c|}{a) \ $-\,0.1501$} 
 & \multicolumn{1}{c|}{$0.0261$} & \multicolumn{1}{|c|}{} &   \multicolumn{1}{|c|}{3} &
  \multicolumn{1}{c|}{a) \ $-\,0.0412$}  &  \multicolumn{1}{c|}{$0.0068$}\\
  
 \multicolumn{1}{|c|}{} &   \multicolumn{1}{|c|}{} & \multicolumn{1}{c|}{b) \ $-\,0.1535$} 
 & \multicolumn{1}{c|}{$0.0262$} & \multicolumn{1}{|c|}{} &   \multicolumn{1}{|c|}{} & 
 \multicolumn{1}{c|}{b) \ $-\,0.0414$}  & \multicolumn{1}{c|}{$0.0069$}\\

 \multicolumn{1}{|c|}{4494} &   \multicolumn{1}{|c|}{4} & \multicolumn{1}{c|}{a) \ $-\,1.6082$}  & 
 \multicolumn{1}{c|}{$0.0497$} &\multicolumn{1}{|c|}{7454} &   \multicolumn{1}{|c|}{4} &
  \multicolumn{1}{c|}{a) \ $-\,0.4231$}  & \multicolumn{1}{c|}{$0.0144$}\\
  
 \multicolumn{1}{|c|}{} &   \multicolumn{1}{|c|}{} & \multicolumn{1}{c|}{b) \ $-\,1.6084$} 
 & \multicolumn{1}{c|}{$0.0497$} &\multicolumn{1}{|c|}{} &   \multicolumn{1}{|c|}{}
  & \multicolumn{1}{c|}{b) \ $-\,0.4232$}  & \multicolumn{1}{c|}{$0.0145$}\\

 \multicolumn{1}{|c|}{} &   \multicolumn{1}{|c|}{5} & \multicolumn{1}{c|}{1) \ $-\,289.60$}  & 
 \multicolumn{1}{c|}{$8.9556$} &\multicolumn{1}{|c|}{} &   \multicolumn{1}{|c|}{5} &
  \multicolumn{1}{c|}{1) \ $-\,178.75$}  & \multicolumn{1}{c|}{$6.0820$}\\
  
 \multicolumn{1}{|c|}{} &   \multicolumn{1}{|c|}{} & \multicolumn{1}{c|}{2) \ $-\,0.0052$} 
 & \multicolumn{1}{c|}{$0.0002$} &\multicolumn{1}{|c|}{} &   \multicolumn{1}{|c|}{}
  & \multicolumn{1}{c|}{2) \ $-\,0.0014$}  & \multicolumn{1}{c|}{$0.00005$}\\

  \hline        
 \end{tabular}
 
 \vskip 2 ex
 
 For models 3 and 4, the values of $W$ and $T$ are given according to variants (a) and (b), and, for model 5, according to variants (1) and (2). The energies are expressed in units of $10^{55}\, J$.
 
\newpage

9. CONCLUSIONS

\vskip 2 ex

Three new models of an EG, which are in good
agreement with modern ideas about the structure of
such galaxies, have been created for solving some
problems of celestial mechanics and astrophysics.
According to these models, an EG together with the
halo is regarded as a two-layer inhomogeneous ellip-
soid of revolution: a prolate spheroid. For a triaxial
ellipsoid, the axis of rotation does not coincide with
the apparent minor axis and the isophotes are misaligned. For an oblate spheroid, in which the apparent
axis coincides with the axis of rotation and the isophote alignment is not violated, is not considered. In
this case, the outer and inner layers are assumed to be
similar and concentric and their centers coincide with
the center of the EG. The LP EG is considered the
inner layer and comprises an inhomogeneous prolate
spheroid with a spheroidal density distribution. In the
LP EG, a BM with an astrophysical law of density dis-
tribution prevails. The outer part comprises an inho-
mogeneous spherical layer with a spherical density
distribution (model 3) or a spheroidal layer with a
spheroidal density distribution (model 4). According
to model 3, the outer layer and the halo of the galaxy
are bounded by a sphere with a radius equal to the
scale radius of the EG, and, according to model 4,
they are bounded by a spheroidal surface with a semi-
major axis equal to the scale radius of the galaxy. It is
assumed that the spherical and spheroidal layers
mainly consist of dark matter (DM) and, depending
on its presence in the central regions of the EG, in
models 3 and 4, two variants are considered: (a) the
main part of DM is outside the LP EG and (b) the DM
content in the inner regions of the EG is comparable
to the BM content. In this case, the matching condi-
tions for the potential at the interface between the LP
SC and the spherical (spheroidal) layer are deter-
mined.

According to model 5, an EG with (variant 1) or
without (variant 2) a halo comprises an inhomoge-
neous ellipsoid of revolution: a prolate spheroid con-
sisting of BM and DM. In model 5, there is no inter-
face between the LP SC and the homeoid; therefore,
the fulfillment of the matching conditions for the
potential is not considered.

Within the three new models of a layered inhomo-
geneous EG, the total gravitational (potential) energy,
rotational kinetic energy, and velocity dispersion at a
distance of the effective radius of the galaxy have been
determined. A new method for determining the aver-
age values of the scale radius of an EG, the density at
its center, and the value of the parameter  corre-
sponding to the effective radius of the galaxy is pro-
posed.

The results obtained have been applied to sixty EGs
and are presented in the form of tables for ten galaxies.

Analysis of the equilibrium and stability of a
dynamic system created on the basis of models 3 and 4
will be carried out by the author separately.

 \vskip 4.0ex

ACKNOWLEDGMENTS
 \vskip 2.0ex

I am grateful to professor B.P. Kondratyev for valuable
advices and remarks.

 \vskip 4.0ex

REFERENCES

 \vskip 2.0ex
 
1.S. A. Gasanov, Astron. Rep. 56, 469 (2012).

2.S. A. Gasanov, Astron. Rep. 58, 167 (2014).

3.S. A. Gasanov, Astron. Rep. 59, 238 (2015).

4.A. V. Zasov, A. S. Saburova, A. V. Khoperskov, and
S.A. Khoperskov, Phys. Usp. 60, 3 (2017).

5.G. Bertin, R. P. Saglia, and M. Stiavelli, Astrophys. J.
384, 423 (1992).

6.M. Oguri, C. E. Rusu, and E. E. Falco, Mon. Not. R.
Astron. Soc. 439, 2494 (2014).

7.G. de Vaucouleurs, A. de Vaucouleurs, H. Corwin,
R.J. Buta, G. Paturel, and P. Fouque, Third Reference
Catalouge of Bright Galaxies (Springer, New York,
1991), Vols. 2, 3.

8.B. P. Kondratyev and L. M. Ozernoi, Sov. Astron. Lett.
5, 37 (1979).

9.B. P. Kondratyev and V. S. Kornoukhov, Mon. Not. R.
Astron. Soc. 478, 3159 (2018).

10.B. P. Kondrat’ev, Cand. Sci. Dissertation (Moscow,
1982).

11.B. P. Kondrat’ev, Theory of Potential. New Methods and
Problems with Solutions (Mir, Moscow, 2007) [in Rus-
sian].

12.E. Hubble, Astrophys. J. 71, 231 (1930).

13.J. F. Navarro, C. S. Frenk, and S. D. M. White, Astro-
phys. J. 490, 493 (1997).

14.P. Cote, D. E. McLaughlin, J. G. Cohen, and J. P. Bla-
keslee, Astrophys. J. 591, 850 (2003).

15.J. Binney and S. Tremaine, Galactic Dynamics (Prince-
ton Univ. Press, Princeton, 2008).

16.W. Dehnen, Mon. Not. R. Astron. Soc.  265, 250
(1993).

17.S. Tremaine, D. O. Richstone, Y.-I. Byun, A. Dressler,
S. M. Faber, C. Grillmair, J. Kormendy, and T.R.Lauer,
Astron. J. 107, 634 (1994).

18.L. Hernquist, Astrophys. J. 356, 359 (1990).

19.W. Jaffe, Mon. Not. R. Astron. Soc. 202, 995 (1983).

20.G. N. Duboshin,  Celestial Mechanics. Fundamental
Problems and Methods (Nauka, Moscow, 1968) [in
Russian].

21.J. P. Ostriker and P. J. E. Peebles, Astrophys. J. 186,
467(1973).

22.V. L. Polyachenko and A. M. Fridman, Physics of Grav-
itating Systems I: Equilibrium and Stability (Nauka,
Moscow, 1976; Springer, Berlin, 1984).

23.J. Binney, Mon. Not. R. Astron. Soc. 190, 873 (1980).

24.J. Binney, Mon. Not. R. Astron. Soc. 190, 421 (1980).

25.Ph. Prugniel and F. Simien, Astron. Astrophys. 321, 111
(1997).

26.B. Terzic and A. W. Graham, arXiv: astro-ph/0506192
(2005).

27.R. L. Davies, E. M. Sadler, and R. F. Peletier, Mon.
Not. R. Astron. Soc. 262, 650 (1993).

28.S. Samurovic, Astron. Astrophys. 470, A132 (2014).

29.M. Cappellari, E. Emsellem, D. Krajnovic, R. M. Mc-
Dermit, et al., Mon. Not. R. Astron. Soc.  413, 813
(2011).

30.M. Cappellari, N. Scott, K. Alatalo, L. Blitz, et al.,
Mon. Not. R. Astron. Soc. 432, 1709 (2013).

31.L. R. Spitler, A. F. Dunkan, J. Strader, J. P. Brodie, and
J. S. Gallagher, Mon. Not. R. Astron. Soc.  385, 361
(2008).

32.D. A. Forbes, L. Sinpertu, G. Savorgnan, A. J. Roma-
nowsky, C. Usher, and J. Brodie, Mon. Not. R. Astron.
Soc. 464, 4611 (2017).

33.E. van Uitert, M. Cacciato, H. Hoekstra, M. Brower,
etal., Mon. Not. R. Astron. Soc. 459, 3251 (2016).

34.D.-W. Kim and G. Fabbiano, Astrophys. J.  812, 127
(2015).

35.D. N. Spergel, R. Bean, O. Dore, M. R. Nolta, et al.,
Astrophys. J. Suppl. 170, 377 (2007).

\end{document}